\documentclass[a4paper, 12pt]{article}
\usepackage{amsmath, amssymb}
\usepackage[left=10mm, top=10mm, right=10mm, bottom=15mm]{geometry}

\DeclareMathOperator{\im}{Im}

\DeclareMathOperator{\const}{const}

\DeclareMathOperator{\Tr}{Tr}
\DeclareMathOperator{\diag}{diag}

\title{Tomography of multimode quantum systems with quadratic Hamiltonians
       and multivariable Hermite polynomials}
\author{V. I. Man'ko \and V. A. Sharapov \and E. V. Shchukin}
\date{}

\begin{document}
\maketitle
\begin{abstract}
The systems with multimode nonstationary Hamiltonians quadratic in position and momentum
operators are reviewed. The tomographic probability distributions (tomograms) for the Fock
states and Gaussian states of the quadratic systems are discussed. The tomograms for the Fock
states are expressed in terms of multivariable Hermite polynomials.  Using the obvious
physical relations some new formulas for multivariable Hermite polynomials are found. Examples
of oscillator and charge moving in electromagnetic field are presented.
\end{abstract}

\section{Introduction}
The tomographic probability distribution was introduced \cite{fp-17-397}, \cite{pr-a-40-2847}
for reconstructing the Wigner function \cite{pr-40-749} of quantum states. The optical
tomography scheme \cite{prl-70-1244} was used in experiments for measuring quantum states of
photons. Recently the symplectic tomography method was suggested \cite{qso-7-615}. The
possibility to describe the quantum state in terms of tomographic probability distribution was
employed to reformulate quantum mechanics \cite{fp-27-801}, \cite{pl-a-213-1} avoiding the
complex probability amplitudes, wave function and density matrix as conceptual ingredients.
The evolution equation for tomograms \cite{fp-27-801}, \cite{pl-a-213-1} and equation for
energy levels in terms of the tomograms \cite{jrlr-18-407} is tomographic counterpart of Moyal
equations \cite{pcps-45-99} written for the Wigner function of quantum system. The tomographic
probability distribution was introduced also for systems with spin degrees of freedom
(\cite{pl-a-229-335}, \cite{zetf-112-796}, \cite{jetp-114-437}, \cite{pr-a-57-671},
\cite{prl-84-802}). The discrete spin degrees of freedom and the continuous ones like a
position were considered in tomographic representation in \cite{jp-a-34-3461}. The tomograms
are used for description of quantum states because they contain the same information on the
state as other characteristics like Wigner function \cite{pr-40-749}, Glauber-Sudarshan
$P$-function \cite{prl-10-84} and \cite{prl-10-277}, Husimi $Q$-function \cite{ppmsj-23-264},
but the tomograms have specific property of standard probability distribution. Important role
play the quantum systems with Hamiltonians which are quadratic in position and momentum
\cite{malkin-manko-dynamical-symmetries-coherent-states-quantum-systems}. The charge moving in
homogeneous electric and magnetic field and linear vibrations of polyatomic molecules are
examples of such systems. The coherent states and Fock states for such systems were studied
using time-dependent integrals of motion linear in position and momentum  in
\cite{pl-a-30-414}. The wave functions of coherent states and Fock states were obtained for
the quadratic systems in explicit form. On the other hand the tomographic description of the
states of the multimode quadratic system has not been presented till now.

The aim of our work is to obtain the tomograms of the specific quantum states of the quadratic
systems. We find the explicit Gaussian tomograms for coherent (and squeezed) multimode
vibrations. Also we show that the tomograms of Fock states are the positive probability
distributions expressed in terms of multivariable Hermite polynomials. The expression for
transition probabilities obvious from physical point of view generate some new formula for the
Hermite polynomials.

The paper is organized as follows. In Section $2$ we present a review of properties of
coherent and Fock states for multimode systems with nonstationary quadratic Hamiltonians. In
Section $3$ we discuss the coherent and Fock states for such systems in the tomographic
probability representation. In Section $4$ we obtain some new formulas for multivariable
Hermite polynomials. The general results are illustrated by two examples in Section $5$ where
a driven parametric oscillator and a charged particle moving in nonstationary electric field
are considered in tomographic probability representation. Concluding remarks on properties of
tomographic probability representation for quadratic quantum systems are presented in Section
$6$. The mathematical properties of symplectic matrices are described in Appendix A and
calculation of the matrix exponent which is used in main text is presented in Appendix B.

\nocite{qso-8-1017}

\section{Coherent and Fock states of systems with quadratic Hamiltonians}

In this section we review the problem of finding linear integrals of motion and constructing
the coherent and Fock states of nonstationary $N$-dimensional systems described by
Schr\"{o}dinger equation:
\begin{equation} \label{E:eqQ}
    i\hbar\frac{\partial{\psi}({\boldsymbol{x}},t)}{\partial{t}} = \widehat{H}(t)\psi({\boldsymbol{x}},t),
\end{equation}
with an arbitrary time-dependent quadratic Hermitean Hamiltonian,
i.e.
\begin{equation}\label{e:quadricH}
    \widehat{H}(t) = \frac{1}{2}\widehat{{\boldsymbol{q}}}^{T}B(t)\widehat{{\boldsymbol{q}}} + {\boldsymbol{c}}^{T}(t)\widehat{{\boldsymbol{q}}}
                   =\frac{1}{2}\sum\limits^{2N}_{\alpha,\beta=1}\widehat{q}_\alpha B_{\alpha\beta}(t)\widehat{q}_\beta +
                   \sum\limits^{2N}_{\gamma=1}c_\gamma(t)\widehat{q}_\gamma,
\end{equation}
where we use $2N$-vector $\widehat{{\boldsymbol{q}}} = \left(\widehat{{\boldsymbol{p}}},
\widehat{{\boldsymbol{x}}}\right)$, $N$-vector
$\widehat{{\boldsymbol{p}}}=(\widehat{p}_1,.....,\widehat{p}_N)$, $N$-vector
$\widehat{{\boldsymbol{x}}}=(\widehat{x}_1,......,\widehat{x}_N)$, and $
\widehat{p}_m=-i\hbar\partial/\partial{x_m}$, $\widehat{x}_m=x_m$ in position representation,
$m=1,... ,N$. The notation $\widehat{\boldsymbol{q}}$ is used for column vector, and the
notation $\widehat{\boldsymbol{q}}^T$ is used for row vector which is transposed column
vector. The matrix elements $B_{\alpha\beta}(t)$ form the $4$-block matrix
\begin{equation}
    B(t) =
    \begin{Vmatrix} B_{pp}(t) & B_{px}(t) \\ \\
                    B_{xp}(t) & B_{xx}(t)
    \end{Vmatrix}
\end{equation}
which is a real $2N\times 2N$-symmetric matrix, and $2N$-vector ${\boldsymbol{c}}(t) =
\left({\boldsymbol{c}}_p(t), {\boldsymbol{c}}_x(t)\right)$ with $N$-vectors
 ${\boldsymbol{c}}_{p}(t)=(c_{p1}(t),.....,c_{pN}(t))$,
 ${\boldsymbol{c}}_{x}(t)=(c_{x1}(t),.....,c_{xN}(t))$
are arbitrary real vectors. Coherent and Fock states of the system
with Hamiltonian \eqref{e:quadricH} can be obtained in the
framework of method of time-dependent invariants
(\cite{malkin-manko-dynamical-symmetries-coherent-states-quantum-systems}).

Let us consider the following operators:
\begin{align}
    \widehat{P}_{m}(t) &= \widehat{U}(t)\widehat{p}_{m}\widehat{U}^{-1}(t), &
    \widehat{Q}_{\alpha}(t) &= \widehat{U}(t)\widehat{q}_{\alpha}\widehat{U}^{-1}(t), \\
    \widehat{X}_{m}(t) &= \widehat{U}(t)\widehat{x}_{m}\widehat{U}^{-1}(t), &
    \widehat{{\boldsymbol{Q}}}(t) &= (\widehat{Q}_{1}(t),......,\widehat{Q}_{2N}(t)) = (\widehat{{\boldsymbol{P}}}(t),\widehat{{\boldsymbol{X}}}(t)).
\end{align}
where $\widehat{U}(t)$ is unitary evolution operator, which connects wave function
$\psi({\boldsymbol{x}},t)$ given at the moment $t$ with the wave function
$\psi({\boldsymbol{x}},0)$ given at the moment $t=0$
\begin{equation}
    \psi({\boldsymbol{x}},t)=\widehat{U}(t)\psi({\boldsymbol{x}},0).
\end{equation}
One can verify that the operators $\widehat{Q}_{\alpha}(t)$
satisfy the following equation
\begin{equation}\label{e:I}
    i\hbar\frac{\partial\widehat{Q}_{\alpha}(t)}{\partial t} = \left[\widehat{H}(t),
    \widehat{Q}_{\alpha}(t)\right].
\end{equation}
It means that the operators $\widehat{Q}_{\alpha}(t)$ are quantum
integrals of motion, i.e. their mean values
${\left<\widehat{Q}_{\alpha}(t)\right>}_\psi$ remain constant on
an arbitrary solution $\psi$ of Sch\"{o}dinger equation
\begin{equation}
    \frac{d}{dt}{\left<\widehat{Q}_{\alpha}(t)\right>}_\psi = 0.
\end{equation}
One can verify that $\widehat{{\boldsymbol{Q}}}(t)$ satisfying
\eqref{e:I}, where $\widehat{H}(t)$ is a quadratic Hamiltonian, is
given by the formula
\begin{equation}\label{e:i}
    \widehat{{\boldsymbol{Q}}}(t) = \Lambda(t)\widehat{{\boldsymbol{q}}} + {\boldsymbol{\Delta}}(t),
\end{equation}
where real $2N\times 2N$-matrix $\Lambda(t)$ and real $2N$-vector ${\boldsymbol{\Delta}}(t)$
read
\begin{equation}
    \Lambda(t) =
    \begin{Vmatrix}
        \Lambda_{pp}(t) & \Lambda_{px}(t) \\ \\
        \Lambda_{xp}(t) & \Lambda_{xx}(t)
    \end{Vmatrix},\qquad\qquad
    {\boldsymbol{\Delta}}(t) =
    \begin{pmatrix}
        {\boldsymbol{\Delta}}_p(t) \\
        {\boldsymbol{\Delta}}_x(t)
    \end{pmatrix}.
\end{equation}
Indeed, substituting this expression into \eqref{e:I}, we have
\begin{align}\label{e:L}
\begin{split}
    i\hbar\dot{\Lambda}(t)\widehat{{\boldsymbol{q}}} + i\hbar\dot{{\boldsymbol{\Delta}}}(t) =
    \left[\frac{1}{2}\widehat{{\boldsymbol{q}}}^TB(t)\widehat{{\boldsymbol{q}}}+{\boldsymbol{c}}^T(t)\widehat{{\boldsymbol{q}}}, \Lambda(t)\widehat{{\boldsymbol{q}}}+{\boldsymbol{\Delta}}(t)\right] =
    \frac{1}{2}\Lambda(t)\left[\widehat{{\boldsymbol{q}}}^TB(t)\widehat{{\boldsymbol{q}}},\widehat{{\boldsymbol{q}}}\right] + \Lambda(t)\left[{\boldsymbol{c}}^T(t)\widehat{{\boldsymbol{q}}},\widehat{{\boldsymbol{q}}}\right],
\end{split}
\end{align}
where we introduce the following notations (
$\widehat{{\boldsymbol{a}}}$ and $\widehat{{\boldsymbol{b}}}$ are
arbitrary $N$-vectors):
\begin{align}
\begin{split}
    \left[\widehat{{\boldsymbol{a}}}, \widehat{{\boldsymbol{b}}}\right] =
    \left[\begin{pmatrix} \widehat{a}_1 \\ \vdots \\ \widehat{a}_N\end{pmatrix},
          \begin{pmatrix} \widehat{b}_1 \\ \vdots \\ \widehat{b}_N\end{pmatrix}\right] =
    \begin{Vmatrix}\left[\widehat{a}_1, \widehat{b}_1\right] & \hdots & \left[\widehat{a}_1, \widehat{b}_N\right] \\
                   \hdotsfor{3}                                                               \\
                   \left[\widehat{a}_N, \widehat{b}_1\right] & \hdots & \left[\widehat{a}_N,\widehat{b}_N\right]\end{Vmatrix},
    \qquad\text{i.e}\ {\left[\widehat{{\boldsymbol{a}}}, \widehat{{\boldsymbol{b}}}\right]}_{ij} = \left[\widehat{a}_i,
    \widehat{b}_j\right].
\end{split}
\end{align}
Using commutation relations for operators
$\widehat{{\boldsymbol{q}}}$:
\begin{equation}
    \left[\widehat{{\boldsymbol{q}}},\widehat{{\boldsymbol{q}}}\right] = -i\hbar\Sigma_{2N} =
    -i\hbar\begin{Vmatrix} 0 & -E_N \\ \\ E_N & 0 \end{Vmatrix},
\end{equation}
we can calculate commutators on the right hand side of equation
\eqref{e:L}. Finally we obtain the equation
\begin{equation}
    \dot{\Lambda}(t)\widehat{{\boldsymbol{q}}} + \dot{{\boldsymbol{\Delta}}}(t) =
    \Lambda(t) \Sigma_{2N}B(t)\widehat{{\boldsymbol{q}}} +
    \Lambda(t)\Sigma_{2N}{\boldsymbol{c}}(t).
\end{equation}
From this equation we can conclude that $\widehat{{\boldsymbol{Q}}}(t)$, given by the
expression \eqref{e:i}, satisfies \eqref{e:I} provided the matrix $\Lambda(t)$ and the vector
$\Delta(t)$ satisfy the following evolution equations
\begin{equation}\label{e:Ld}
\begin{split}
    \dot{\Lambda}(t) &= \Lambda(t)\Sigma_{2N}B(t), \\
    \dot{{\boldsymbol{\Delta}}}(t)  &= \Lambda(t)\Sigma_{2N}{\boldsymbol{c}}(t).
\end{split}
\end{equation}
The initial conditions for these evolution equations are taken in the form
\begin{equation}\label{e:ini}
    \Lambda(0)=E_{2N}, \qquad\qquad   {\boldsymbol{\Delta}}(0)=0.
\end{equation}
The above initial conditions correspond to initial values of the integrals of motion
$\widehat{\boldsymbol{Q}}(0) = \widehat{\boldsymbol{q}}$. The coherent states of the quadratic
multimode systems play the special role. These states have the Gaussian form and they are
close to classical states of the vibrating oscillators. The states are labeled by continuous
complex quantum numbers. The coherent states form the overcomplete nonorthogonal basis in the
Hilbert space. To find the coherent states of the system with the Hamiltonian
\eqref{e:quadricH}, let us introduce the annihilation and creation operators
\begin{equation}\label{e:a1}
    \widehat{{\boldsymbol{a}}} = A_p\widehat{{\boldsymbol{p}}} + A_x\widehat{{\boldsymbol{x}}}, \qquad \qquad
    \widehat{{\boldsymbol{a}}}^+ = A_p^*\widehat{{\boldsymbol{p}}} + A_x^*\widehat{{\boldsymbol{x}}}, \qquad \qquad
\end{equation}
where $A_{p}$ and $A_{x}$ are time-independent $N\times N$-matrix, and $2N$ annihilation and
creation operators $\widehat{{\boldsymbol{a}}}$ and $\widehat{{\boldsymbol{a}}}^+$ satisfy the
following commutation relations:
\begin{equation}\label{e:a2}
    \left[\begin{pmatrix}
              \widehat{\boldsymbol{a}} \\ \widehat{\boldsymbol{a}}^+
          \end{pmatrix},
          \begin{pmatrix}
              \widehat{\boldsymbol{a}} \\ \widehat{\boldsymbol{a}}^+
          \end{pmatrix}\right] =
          \Sigma_{2N}.
\end{equation}
To satisfy the condition \eqref{e:a2}, the matrices $A_{p}$ and $A_{x}$ must possess the
following properties:
\begin{align}
    A_xA^T_p - A_pA^T_x &= 0,\label{e:A1} \\
    A_xA^+_p - A_pA^+_x &= -\frac{i}{\hbar}E_N.\label{e:A2}
\end{align}
In Appendix A we show that such matrices also possess the
following properties
\begin{enumerate}
    \renewcommand{\theenumi}{(\roman{enumi})}
    \renewcommand{\labelenumi}{\theenumi}

    \item
    Matrices $A_p$ and $A_x$ are non-singular,

    \item
    $A^T_pA^*_p = A^+_pA_p,\quad A^T_xA^*_x=A^+_xA_x$,

    \item
    $A^+_xA_p-A^T_xA^*_p = A^T_pA^*_x-A^+_pA_x =
    \displaystyle\frac{i}{\hbar}E_N$.
\end{enumerate}
Below we use this properties to transform some expressions. Making transformation $
\widehat{a}_{k} \rightarrow \widehat{A}_{k}(t)=
\widehat{U}(t)\widehat{a}_{k}\widehat{U}(t)^{-1}$ on both sides of \eqref{e:a1} we obtain
integrals of motion in explicit form
\begin{equation}
    \widehat{{\boldsymbol{A}}}(t) = A\widehat{{\boldsymbol{Q}}}(t) = \Omega(t)\widehat{{\boldsymbol{q}}}+{\boldsymbol{\delta}}(t),\qquad
    \Omega(t) = A\Lambda(t), \qquad
    {\boldsymbol{\delta}}(t) = A{\boldsymbol{\Delta}}(t),
\end{equation}
where the rectangular matrix
\begin{equation}
    A=\left\| A_p, A_x \right\|
\end{equation}
contains time-independent $N \times N$-blocks and the $N$-vector
\begin{equation}
    \widehat{{\boldsymbol{A}}}(t) = \left(\widehat{A}_1(t),\ldots,\widehat{A}_N(t)\right)
\end{equation}
takes the initial value
\begin{equation}
    \widehat{{\boldsymbol{A}}}(0) = \widehat{{\boldsymbol{a}}}.
\end{equation}
Making the same procedure as we make obtaining the equations for the matrix $\Lambda(t)$ we
obtain the equations for the matrix $\Omega(t)$ and vector ${\boldsymbol{\delta}}(t)$ that are
similar to \eqref{e:Ld}
\begin{equation}\label{e:LD}
\begin{split}
    \dot{\Omega}(t) &= \Omega(t)\Sigma_{2N}B(t), \\
    \dot{{\boldsymbol{\delta}}}(t)  &= \Omega(t)\Sigma_{2N}{\boldsymbol{c}}(t).
\end{split}
\end{equation}
But the initial conditions for the matrix $\Omega(t)$ and the vector $\boldsymbol{\delta}(t)$
are different from \eqref{e:ini}, i.e.
\begin{equation}
    \Omega(0)=A, \qquad\qquad {\boldsymbol{\delta}}(0)=0.
\end{equation}
Let us introduce the following matrices:
\begin{align}
\begin{split}
    \Lambda_p(t) &= A_p\Lambda_{pp}(t) + A_x\Lambda_{xp}(t), \\
    \Lambda_x(t) &= A_p\Lambda_{px}(t) + A_x\Lambda_{xx}(t).
\end{split}
\end{align}
The matrices $\Lambda_p(t)$ and $\Lambda_x(t)$ possess the properties similar to corresponding
ones for the matrices $A_p$ and $A_x$ (see Appendix A). Using the matrices we can express the
integrals of motion $\widehat{{\boldsymbol{A}}}(t)$ through the momentum and position
operators $\widehat{{\boldsymbol{p}}}$ and $\widehat{{\boldsymbol{x}}}$:
\begin{align}\label{e:A}
    \widehat{{\boldsymbol{A}}}(t)&=\Lambda_p(t)\widehat{{\boldsymbol{p}}} + \Lambda_x(t)\widehat{{\boldsymbol{x}}} + {\boldsymbol{\delta}}(t), \\
    \Omega(t) &= \left\| \Lambda_p(t),\Lambda_{x}(t) \right\|.
\end{align}

Let us introduce the definition of coherent and Fock states of the system with Hamiltonian
$\widehat{H}(t)$. For all ${\boldsymbol{\alpha}} =
\left(\alpha_1,\ldots,\alpha_N\right)\in\mathbf{C}^N$ there exists the normalised
eigenfunction $\psi_{{\boldsymbol{\alpha}}}({\boldsymbol{x}},t)$ of operators
$\widehat{A}_k(t)$
\begin{equation}
    \widehat{A}_k(t)\psi_{{\boldsymbol{\alpha}}}({\boldsymbol{x}},t) = \alpha_k\psi_{{\boldsymbol{\alpha}}}({\boldsymbol{x}},t),
\end{equation}
which is called the wave function of the coherent state; and for
all ${\boldsymbol{n}} =
\left(n_1,\ldots,n_N\right)\in\mathbf{N}^N$ there exists the
normalised eigenfunction
$\psi_{{\boldsymbol{n}}}({\boldsymbol{x}},t)$ of operators
$\widehat{A}_{k}^{+}(t)\widehat{A}_{k}(t)$
\begin{equation}
    \widehat{A}_{k}^{+}(t)\widehat{A}_{k}(t)\psi_{{\boldsymbol{n}}}({\boldsymbol{x}},t)=n_{k}\psi_{{\boldsymbol{n}}}({\boldsymbol{x}},t),
\end{equation}
which is called the wave function of the Fock state. It is well known
\cite{malkin-manko-dynamical-symmetries-coherent-states-quantum-systems} that we can represent
the coherent state in the form of series:
\begin{equation}\label{e:psialph}
   \psi_{{\boldsymbol{\alpha}}}({\boldsymbol{x}},t) = e^{-\frac{1}{2}{\Vert{\boldsymbol{\alpha}}\Vert}^2}
   \sum_{{\boldsymbol{m}}={\boldsymbol{0}}}^{\infty}\psi_{{\boldsymbol{m}}}({\boldsymbol{x}},t)\frac{{{\boldsymbol{\alpha}}}^{{\boldsymbol{m}}}}{\sqrt{{\boldsymbol{m}}!}}.
\end{equation}
Here we use the following notations:
\begin{equation}
    {\boldsymbol{x}}! = \prod^N_{i=1}x_i!,\qquad
    {\boldsymbol{x}}^{{\boldsymbol{k}}} = \prod^N_{i=1}x^{k_i}_i,\qquad
    {\Vert{\boldsymbol{x}}\Vert}^2 = \sum^N_{i=1}{\left\vert x_i\right\vert}^2.
\end{equation}
It is easy to show that for all ${\boldsymbol{\alpha}} =
\left(\alpha_1,\ldots,\alpha_N\right)\in\mathbf{C}^N$ the function
$f_{{\boldsymbol{\alpha}}}({\boldsymbol{x}},t)$, defined as
\begin{equation}
     f_{{\boldsymbol{\alpha}}}({\boldsymbol{x}},t) = \psi_{{\boldsymbol{\alpha}}}({\boldsymbol{x}},t)e^{\frac{1}{2}{\Vert{\boldsymbol{\alpha}}\Vert}^2},
\end{equation}
is an analytic function of complex argument ${\boldsymbol{\alpha}}$ and it satisfies the
following equations:
\begin{equation}\label{e:alpha1}
    \widehat{A}_k(t)f_{{\boldsymbol{\alpha}}}({\boldsymbol{x}},t) = \alpha_k f_{{\boldsymbol{\alpha}}}({\boldsymbol{x}},t),  \qquad \qquad
    \widehat{A}^+_k(t)f_{{\boldsymbol{\alpha}}}({\boldsymbol{x}},t) = \frac{\partial f_{{\boldsymbol{\alpha}}}({\boldsymbol{x}},t)}{\partial\alpha_k}.
\end{equation}
Indeed, we can represent the function $f_{{\boldsymbol{\alpha}}}({\boldsymbol{x}},t)$ in the
form of the series
\begin{equation} \label{e:alpha2}
    f_{{\boldsymbol{\alpha}}}({\boldsymbol{x}},t) = \psi_{{\boldsymbol{\alpha}}}({\boldsymbol{x}},t)e^{\frac{1}{2}{\Vert{\boldsymbol{\alpha}}\Vert}^2} =
    \sum_{{\boldsymbol{m}}={\boldsymbol{0}}}^{\infty}\psi_{{\boldsymbol{m}}}({\boldsymbol{x}},t)\frac{{{\boldsymbol{\alpha}}}^{{\boldsymbol{m}}}}{\sqrt{{\boldsymbol{m}}!}}.
\end{equation}
It follows from this expression that the function
$f_{{\boldsymbol{\alpha}}}({\boldsymbol{x}},t)$ is analytic function. To prove the validity of
the action of the operators $A_m(t)$ and $A^+_m(t)$ on the function
$f_{\boldsymbol{\alpha}}(\boldsymbol{x},t)$ we have to check \eqref{e:alpha1} at the time
moment $t=0$, i.e. to check the following equations:
\begin{equation}\label{e:0}
    \widehat{a}_k f_{{\boldsymbol{\alpha}}}({\boldsymbol{x}},0) = \alpha_k f_{{\boldsymbol{\alpha}}}({\boldsymbol{x}},0),  \qquad \qquad
    \widehat{a}^+_k f_{{\boldsymbol{\alpha}}}({\boldsymbol{x}},0) = \frac{\partial f_{{\boldsymbol{\alpha}}}({\boldsymbol{x}},0)}{\partial\alpha_k}.
\end{equation}
The first equation of \eqref{e:0} follows from the definition of
the function $f_{{\boldsymbol{\alpha}}}({\boldsymbol{x}},t)$. To
check the second equation of \eqref{e:0} we act on the function
$f_{{\boldsymbol{\alpha}}}({\boldsymbol{x}},0)$ by operator
$\widehat{a}^+_k$:
\begin{equation}\label{e-37}
    \widehat{a}^+_kf_{\boldsymbol{\alpha}}({\boldsymbol{x}},0) = \widehat{a}^+_k\left(\sum_{{\boldsymbol{m}}={\boldsymbol{0}}}^{\infty}\frac{m_k\psi_{{\boldsymbol{m}}}({\boldsymbol{x}},0)}{\sqrt{{\boldsymbol{m}}!}}{{\boldsymbol{\alpha}}}^{{\boldsymbol{m}}}\right)
    = \sum_{{\boldsymbol{m}}={\boldsymbol{0}}}^{\infty}\frac{\sqrt{m_k+1}\psi_{{\boldsymbol{m}}+{\boldsymbol{e}}_k}({\boldsymbol{x}},0)}{\sqrt{{\boldsymbol{m}}!}}{{\boldsymbol{\alpha}}}^{{\boldsymbol{m}}}
\end{equation}
and differentiate it with respect to variable $\alpha_k$
\begin{align}
\begin{split}\label{e-38}
    \frac{\partial f_{\boldsymbol{\alpha}}(\boldsymbol{x},0)}{\partial\alpha_k} &= \frac{\partial}{\partial\alpha_k}\left(
    \sum_{{\boldsymbol{m}}={\boldsymbol{0}}}^{\infty}\frac{\psi_{{\boldsymbol{m}}}({\boldsymbol{x}},0)}{\sqrt{{\boldsymbol{m}}!}}{{\boldsymbol{\alpha}}}^{{\boldsymbol{m}}}\right)
    = \sum_{{\boldsymbol{m}}={\boldsymbol{0}}}^{\infty}\frac{m_k\psi_{{\boldsymbol{m}}}({\boldsymbol{x}},0)}{\sqrt{{\boldsymbol{m}}!}}{{\boldsymbol{\alpha}}}^{{\boldsymbol{m}}-{\boldsymbol{e}}_k}
    = \sum_{{\boldsymbol{m}}={\boldsymbol{0}}}^{\infty}\frac{\sqrt{m_k}\psi_{{\boldsymbol{m}}}({\boldsymbol{x}},0)}{\sqrt{\left({\boldsymbol{m}}-{\boldsymbol{e}}_k\right)!}}{{\boldsymbol{\alpha}}}^{{\boldsymbol{m}}-{\boldsymbol{e}}_k}\\
    &= \sum_{{\boldsymbol{m}}={\boldsymbol{0}}}^{\infty}\frac{\sqrt{m_k+1}\psi_{{\boldsymbol{m}}+{\boldsymbol{e}}_k}({\boldsymbol{x}},0)}{\sqrt{{\boldsymbol{m}}!}}{{\boldsymbol{\alpha}}}^{{\boldsymbol{m}}}
\end{split}
\end{align}
where
\begin{equation}
    {\boldsymbol{e}}_k = (\overbrace{\underbrace{0,\ldots,0,1}_k,0,\ldots,0}^N)
\end{equation}
(unit on the $k$-th position). Comparing the two relations \eqref{e-37} and \eqref{e-38} we
can conclude that the function $f_{{\boldsymbol{\alpha}}}({\boldsymbol{x}},0)$ satisfies
\eqref{e:0}. Making some transformations on both sides of \eqref{e:0} by means of the
evolution operator
\begin{align}
    \widehat{U}(t)\widehat{a}_k \widehat{U}^{-1}(t)\widehat{U}(t)f_{{\boldsymbol{\alpha}}}({\boldsymbol{x}},0) = \widehat{U}(t)\alpha_k f_{{\boldsymbol{\alpha}}}({\boldsymbol{x}},0),\\
    \widehat{U}(t) \widehat{A}^+_k \widehat{U}^{-1}(t)\widehat{U}(t)f_{{\boldsymbol{\alpha}}}({\boldsymbol{x}},0) = \widehat{U}(t)\frac{\partial f_{{\boldsymbol{\alpha}}}({\boldsymbol{x}},0)}{\partial\alpha_k},
\end{align}
and using the relation
$f_{{\boldsymbol{\alpha}}}({\boldsymbol{x}},t)=\widehat{U}(t)f_{{\boldsymbol{\alpha}}}({\boldsymbol{x}},0)$
we obtain \eqref{e:alpha1}. Therefore if the function
$f_{{\boldsymbol{\alpha}}}({\boldsymbol{x}},0)$ satisfies \eqref{e:0} the function
$f_{{\boldsymbol{\alpha}}}({\boldsymbol{x}},t)$ satisfies \eqref{e:alpha1}. Substituting the
expression \eqref{e:A} for operator ${\boldsymbol{\widehat{A}}}(t)$ into the first equation of
\eqref{e:alpha1} we obtain
\begin{equation}\label{e:ieq}
    \widehat{{\boldsymbol{A}}}(t)f_{\boldsymbol{\alpha}}({\boldsymbol{x}},t) =
    -i\hbar\Lambda_p(t)\frac{\partial f_{\boldsymbol{\alpha}}({\boldsymbol{x}},t)}{\partial{\boldsymbol{x}}}+
    \Lambda_x(t){\boldsymbol{x}}f({\boldsymbol{x}},t) + {\boldsymbol{\delta}}(t)f_{\boldsymbol{\alpha}}({\boldsymbol{x}},t) =
    {\boldsymbol{\alpha}}f_{\boldsymbol{\alpha}}({\boldsymbol{x}},t).
\end{equation}
We can rewrite this equation in the form
\begin{equation}\label{e:dfdx}
    -i\hbar\frac{\partial f_{\boldsymbol{\alpha}}({\boldsymbol{x}},t)}{\partial{\boldsymbol{x}}} =
    \Lambda^{-1}_p(t)\Bigl({\boldsymbol{\alpha}}-{\boldsymbol{\delta}}(t) -
    \Lambda_x(t){\boldsymbol{x}}\Bigr)f_{\boldsymbol{\alpha}}({\boldsymbol{x}},t).
\end{equation}
It is easy to show that solution of equation \eqref{e:dfdx} is given by the Gaussian function
\begin{equation}
    f_{\boldsymbol{\alpha}}({\boldsymbol{x}},t) = f_1({\boldsymbol{\alpha}},t)\exp\Biggl\{-\frac{i}{2\hbar}{\boldsymbol{x}}^T\Lambda^{-1}_p(t)\Lambda_x(t){\boldsymbol{x}}
    +\frac{i}{\hbar}{\boldsymbol{x}}^T\Lambda^{-1}_p(t)({\boldsymbol{\alpha}}-{\boldsymbol{\delta}}(t))\Biggr\}.
\end{equation}
To find the function $f_1({\boldsymbol{\alpha}},t)$ we substitute
this expression into the second equation of \eqref{e:alpha1}
\begin{equation}
    \Lambda^*_p(t)\left(-i\hbar\frac{\partial f_{\boldsymbol{\alpha}}({\boldsymbol{x}},t)}{\partial{\boldsymbol{x}}}\right)+
    \Lambda^*_x(t){\boldsymbol{x}}f_1({\boldsymbol{\alpha}},t) + {\boldsymbol{\delta}}^{*}(t)f_1({\boldsymbol{\alpha}},t) =
    \frac{\partial f_1({\boldsymbol{\alpha}},t)}{\partial\boldsymbol{\alpha}} + \frac{i}{\hbar}{\left(\Lambda^T_p(t)\right)}^{-1}{\boldsymbol{x}}f_1({\boldsymbol{\alpha}},t)
\end{equation}
Substituting the expression for $ -i\hbar\frac{\partial
f_{\boldsymbol{\alpha}}}{\partial{\boldsymbol{x}}}$ from equation \eqref{e:dfdx} we get the
equation
\begin{equation}
    \Lambda^*_p(t)\Lambda^{-1}_p(t)\Bigl({\boldsymbol{\alpha}}-{\boldsymbol{\delta}}(t)-\Lambda_x(t){\boldsymbol{x}}\Bigr)f_1({\boldsymbol{\alpha}},t) +
    \Lambda^*_x(t){\boldsymbol{x}}f_1({\boldsymbol{\alpha}},t) = \frac{\partial f_1({\boldsymbol{\alpha}},t)}{\partial{\boldsymbol{\alpha}}} + \frac{i}{\hbar}{\left(\Lambda^T_p(t)\right)}^{-1}{\boldsymbol{x}}f_1({\boldsymbol{\alpha}},t),
\end{equation}
or
\begin{equation}\label{e:f0}
    \frac{\partial f_1({\boldsymbol{\alpha}},t)}{\partial{\boldsymbol{\alpha}}} = \Lambda^*_p(t)\Lambda^{-1}_p(t)({\boldsymbol{\alpha}}-{\boldsymbol{\delta}}(t))f_1({\boldsymbol{\alpha}},t) +
    \left(\Lambda^*_x(t)-\Lambda^*_p(t)\Lambda^{-1}_p(t)\Lambda_x(t) - \frac{i}{\hbar}{\left(\Lambda^T_p(t)\right)}^{-1}\right){\boldsymbol{x}}f_1({\boldsymbol{\alpha}},t).
\end{equation}
Using the properties of matrices $\Lambda_{x}(t)$ and $\Lambda_{p}(t)$ from Appendix A one can
show that the matrix in front of the ${\boldsymbol{x}}$ is equal to zero which can be inferred
from the equality
\begin{equation}
    \Lambda^T_p(t)\Lambda^*_x(t)-\Lambda^T_p(t)\Lambda^*_p(t)\Lambda^{-1}_p(t)\Lambda_x(t)
     = \Lambda^T_p(t)\Lambda^*_x(t)-\Lambda^+_p(t)\Lambda_x(t) = \frac{i}{\hbar}E_N.
\end{equation}
Solving equation \eqref{e:f0} we obtain the expression for the function
$f_1({\boldsymbol{\alpha}},t)$:
\begin{equation}
    f_1({\boldsymbol{\alpha}},t) = \exp\Biggl\{\frac{1}{2}{\boldsymbol{\alpha}}^T\Lambda^*_p(t)\Lambda^{-1}_p(t){\boldsymbol{\alpha}} +
    {\boldsymbol{\alpha}}^T\Bigl({\boldsymbol{\delta}}^{*}(t) - \Lambda_p^{*}(t)\Lambda_p^{-1}(t){\boldsymbol{\delta}}(t)\Bigr)+ \varphi(t)\Biggr\}
\end{equation}
which contains the unknown function of time $\varphi(t)$. Therefore, the wave function of
multimode coherent state reads
\begin{equation}\label{e:ps}
\begin{split}
    \psi_{{\boldsymbol{\alpha}}}({\boldsymbol{x}},t) &= \exp\Biggl\{ -\frac{i}{2\hbar}{\boldsymbol{x}}^T\Lambda^{-1}_p(t)\Lambda_x(t){\boldsymbol{x}} +
    \frac{i}{\hbar}{\boldsymbol{x}}^T\Lambda^{-1}_p(t)({\boldsymbol{\alpha}}-{\boldsymbol{\delta}}(t)) +
    \frac{1}{2}{\boldsymbol{\alpha}}^T\Lambda^*_p(t)\Lambda^{-1}_p(t){\boldsymbol{\alpha}}\\
    &+\boldsymbol{\alpha}^T\Bigl(\boldsymbol{\delta}^*(t)-\Lambda^*_p(t)\Lambda^{-1}_p(t)\boldsymbol{\delta}(t)\Bigr)-\frac{1}{2}{\Vert{\boldsymbol{\alpha}}\Vert}^2 + \varphi(t) \Biggr\}.
\end{split}
\end{equation}
To find the explicit dependence on time of the function $\varphi(t)$ we substitute the
expression \eqref{e:ps} into the Schr\"{o}dinger equation \eqref{E:eqQ}. Using the following
equalities ($A$ and $C$ are arbitrary symmetric matrices):
\begin{align}
\begin{split}
    \widehat{{\boldsymbol{p}}}C\widehat{{\boldsymbol{p}}}({\boldsymbol{x}}A{\boldsymbol{x}}) &= -2\hbar^2\Tr(AC),\\
    \widehat{{\boldsymbol{p}}}({\boldsymbol{x}}A{\boldsymbol{x}}) &= -2i\hbar A{\boldsymbol{x}}.
\end{split}
\end{align}
we obtain the equation for the unknown function $\varphi(t)$:
\begin{equation}\label{e:Sp}
    \frac{\partial{\varphi}(t)}{\partial{t}} = \frac{1}{2}
    \Tr\Bigl(\Lambda_p^{-1}(t)\Lambda_x(t)B_{pp}(t) - B_{xp}(t)\Bigr) +
    \frac{i}{\hbar}{\boldsymbol{c}}_p^T(t)\Lambda_p^{-1}(t){\boldsymbol{\delta}}(t)-
    \frac{i}{2\hbar}{\boldsymbol{\delta}}^T(t)({\Lambda_p^{T}(t)})^{-1}B_{pp}(t)
    \Lambda_p^{-1}(t){\boldsymbol{\delta}}(t).
\end{equation}
In order to solve this equation let us rewrite equations \eqref{e:Ld} in more detail, i.e. we
get the evolution equations for the matrices
\begin{align}\label{e:ur1}
    \dot{\Lambda}_p(t) &= \Lambda_p(t)B_{xp}(t) - \Lambda_x(t)B_{pp}(t), & \dot{\Lambda}_p^{*}(t) &= \Lambda_p^*(t)B_{xp}(t) - \Lambda_x^*(t)B_{pp}(t),\\
    \dot{\Lambda}_x(t) &= \Lambda_p(t)B_{xx}(t) - \Lambda_x(t)B_{px}(t), & \dot{\Lambda}_x^{*}(t) &= \Lambda_p^*(t)B_{xx}(t) - \Lambda_x^*(t)B_{px}(t),
\end{align}
and we get the evolution equations for the vectors
\begin{equation}
    \dot{{\boldsymbol{\delta}}}(t) = \Lambda_p(t){\boldsymbol{c}}_x(t) - \Lambda_x(t){\boldsymbol{c}}_p(t),
    \dot{{\boldsymbol{\delta}}}^*(t) = \Lambda_p^*(t){\boldsymbol{c}}_x(t) - \Lambda_x^*(t){\boldsymbol{c}}_p(t).
\end{equation}
Using the evolution equation for the matrices \eqref{e:ur1} we can transform the expression
for which we calculate the trace of the matrix
\begin{equation}
    \Lambda_p^{-1}(t)\Lambda_x(t)B_{pp}(t) - B_{xp}(t) =
    -\Lambda_p^{-1}(t)\dot{\Lambda}_p(t)
\end{equation}
Taking into account the known formula
\begin{equation}
    \Tr\Bigl(A^{-1}(t)\dot{A}(t)\Bigr)=\frac{d}{dt}\Bigl(\ln{\det A(t)}\Bigr),
\end{equation}
we obtain the relation
\begin{equation}
    \Tr\Bigl(\Lambda_p^{-1}(t)\Lambda_x(t)B_{pp}(t) - B_{xp}(t)\Bigr) = -\frac{d}{dt}\Bigl(\ln{\det\Lambda_p(t)}\Bigr).
\end{equation}
To transform other terms on the right hand side of \eqref{e:Sp} we exclude the vector
${\boldsymbol{c}}_p(t)$ and the matrix $B_{pp}(t)$ by means of \eqref{e:Ld}, i.e.
\begin{equation}\label{e:zam}
    B_{pp}(t)=i\hbar\Bigl(\Lambda_p^T(t)\dot{\Lambda}_p^*(t)-\Lambda_p^+(t)\dot{\Lambda}_p(t)\Bigr), \qquad
    {\boldsymbol{c}}_p(t)=i\hbar\Bigl((\dot{{\boldsymbol{\delta}}}^*(t))^T\Lambda_p^T(t)-\Lambda_p^+(t)\dot{{\boldsymbol{\delta}}}(t)\Bigr).
\end{equation}
Substituting these expressions for the vector ${\boldsymbol{c}}_p(t)$ and the matrix
$B_{pp}(t)$ into the right hand side of equation \eqref{e:Sp} we obtain (thereafter we
sometimes omit explicit dependence of some functions either on all or part of their
arguments):
\begin{align}
\begin{split}
    &\frac{i}{\hbar}{\boldsymbol{c}}_p^T \Lambda_p^{-1}{\boldsymbol{\delta}}-
    \frac{i}{2\hbar}{\boldsymbol{\delta}}^T(\Lambda_p^{T})^{-1}B_{pp}
    \Lambda_p^{-1}{\boldsymbol{\delta}}=
    \Bigl(\dot{{\boldsymbol{\delta}}}^T\Lambda_p^* - (\dot{{\boldsymbol{\delta}}}^*)^T\Lambda_p\Bigr)
    \Lambda_p^{-1}{\boldsymbol{\delta}} \\
    &+\frac{1}{2}{\boldsymbol{\delta}}^T(\Lambda_p^T)^{-1}\Bigl(\Lambda_p^T
    \dot{\Lambda}_p^* - \Lambda_p^+
    \dot{\Lambda}_p\Bigr)\Lambda_p^{-1}{\boldsymbol{\delta}}=\dot{{\boldsymbol{\delta}}}^T\Lambda_p^*
    \Lambda_p^{-1}{\boldsymbol{\delta}} - (\dot{{\boldsymbol{\delta}}}^*)^T {\boldsymbol{\delta}}\\
    &+ \frac{1}{2}{\boldsymbol{\delta}}^T\Bigl(\dot{\Lambda}_p^* \Lambda_p^{-1}
    -\Lambda_p^* \Lambda_p^{-1}\dot{\Lambda}_p
    \Lambda_p^{-1}\Bigr){\boldsymbol{\delta}}
    =\frac{1}{2}\frac{d}{dt}\Bigl({\boldsymbol{\delta}}^T
    \Lambda_p^* \Lambda_p^{-1}{\boldsymbol{\delta}}\Bigr)- \frac{1}{2}\frac{d}{dt}
    \Bigl({\boldsymbol{\delta}}^T {\boldsymbol{\delta}}^*\Bigr) +
    \frac{1}{2}\Bigl(\dot{{\boldsymbol{\delta}}}^T
    {\boldsymbol{\delta}}^* - (\dot{{\boldsymbol{\delta}}}^*)^T {\boldsymbol{\delta}}\Bigr).
\end{split}
\end{align}
This expression gives the possibility to solve the equation \eqref{e:Sp}. Finally, we obtain
the following expression for the normalized wave function of the coherent state
\begin{align}\label{e:coh}
\begin{split}
    \psi_{{\boldsymbol{\alpha}}}({\boldsymbol{x}},t) &= {\left(2\pi\hbar^2\right)}^{-N/4}{\left(\det
    \Lambda_p\right)}^{-1/2}\exp\Biggl\{
    -\frac{i}{2\hbar}{\boldsymbol{x}}^T\Lambda^{-1}_p\Lambda_x{\boldsymbol{x}}+
    \frac{i}{\hbar}{\boldsymbol{x}}^T\Lambda^{-1}_p({\boldsymbol{\alpha}}-\boldsymbol{\delta})
    +\frac{1}{2}{\boldsymbol{\alpha}}^T\Lambda^*_p\Lambda^{-1}_p{\boldsymbol{\alpha}}\Biggr.\\
    \Biggl.&+ \boldsymbol{\alpha}^T\Bigl(\boldsymbol{\delta}^*-\Lambda^*_p\Lambda^{-1}_p\boldsymbol{\delta}\Bigr)-
    \frac{1}{2}{\Vert{\boldsymbol{\alpha}}\Vert}^2 + \frac{1}{2}{\boldsymbol{\delta}}^{T}\Lambda_p^{*}
    \Lambda_p^{-1}{\boldsymbol{\delta}} - \frac{1}{2}{\Vert{\boldsymbol{\delta}}\Vert}^{2}+
    i \int\limits_{0}^{t}\ \im\left({\boldsymbol{\dot{\delta}}}^T{\boldsymbol{\delta}}^{*}\right) \,d\tau \Biggr\}
\end{split}
\end{align}
We verify that the function $\psi_{{\boldsymbol{\alpha}}}({\boldsymbol{x}},t)$ is normalized
in the partial case of the Hamiltonian without linear terms, i.e. we take the vector
${\boldsymbol{\delta}} = 0$ and calculate the integral
\begin{align}\label{e:psialpha}
\begin{split}
    I &= \int\ {\vert\psi_{{\boldsymbol{\alpha}}}({\boldsymbol{x}},t)\vert}^2 \,d{\boldsymbol{x}} = {\vert\Psi(t)\vert}^2
    \int\exp\Biggl\{-\frac{i}{2\hbar}{\boldsymbol{x}}^T\left(\Lambda^{-1}_p\Lambda_x-
    {\left(\Lambda^*_p\right)}^{-1}\Lambda^*_x\right){\boldsymbol{x}}\Biggr.\\
    \Biggl.&+\frac{i}{\hbar}{\boldsymbol{x}}^T\left(\Lambda^{-1}_p{\boldsymbol{\alpha}}-
    {\left(\Lambda^*_p\right)}^{-1}{\boldsymbol{\alpha}}^*\right)+
    \frac{1}{2}{\boldsymbol{\alpha}}^T
    \Lambda^*_p\Lambda^{-1}_p{\boldsymbol{\alpha}} + \frac{1}{2}({\boldsymbol{\alpha}}^*)^T
    \Lambda_p{\left(\Lambda^*_p\right)}^{-1}{\boldsymbol{\alpha}}^*-{\Vert{\boldsymbol{\alpha}}\Vert}^2\Biggr\}\,d{\boldsymbol{x}}
\end{split}
\end{align}
where
\begin{equation}
  \Psi(t) = {\left(2\pi\hbar^2\right)}^{-N/4}{\left(\det\Lambda_p(t)\right)}^{-1/2}.
\end{equation}
To evaluate this integral, we transform the matrix of quadratic form in the exponent, using
the properties of matrices $\Lambda_{x}$ and $\Lambda_{p}$, i.e.
\begin{align}\label{e:lam}
\begin{split}
    \Lambda^{-1}_p\Lambda_x-{\left(\Lambda^*_p\right)}^{-1}\Lambda^*_x &=
    \Lambda^{-1}_p\Lambda_x-\Lambda^+_x{\left(\Lambda^+_p\right)}^{-1}
    =\Lambda^{-1}_p\left(\Lambda_x\Lambda^+_p-\Lambda_p\Lambda^+_x\right)
    {\left(\Lambda^+_p\right)}^{-1} =
    -\frac{i}{\hbar}{\left(\Lambda^+_p\Lambda_p\right)}^{-1}.
\end{split}
\end{align}
Substituting \eqref{e:lam} into \eqref{e:psialpha} and evaluating the obtained integral we get
$I=1$. Thus the correct normalization of the wave function of the coherent state is proved.
Using the formula for generating function for multivariable Hermite polynomials
\begin{equation}\label{e:expH}
    \exp{\left(-\frac{1}{2}{\boldsymbol{a}}^TR{\boldsymbol{a}} + {\boldsymbol{a}}^TR{\boldsymbol{x}}\right)} =
    \sum_{{\boldsymbol{m}}={\boldsymbol{0}}}^{\infty}\frac{H_{{\boldsymbol{m}}}^{\{R\}}({\boldsymbol{x}})}
    {\sqrt{{\boldsymbol{m}}!}}{{\boldsymbol{a}}}^{{\boldsymbol{m}}},
\end{equation}
we can represent the coherent state \eqref{e:coh} in the form \eqref{e:psialph}. The
coefficients in front of ${\boldsymbol{a}}^{{\boldsymbol{m}}}$ in \eqref{e:psialph} turn out
to be the multivariable Hermite polynomials. Comparing the coefficients in \eqref{e:expH} with
coefficients in \eqref{e:psialph} we obtain the wave function of the Fock state
\begin{equation}\label{e:psin}
    \psi_{{\boldsymbol{n}}}({\boldsymbol{x}},t)=\frac{1}{\sqrt{{\boldsymbol{n}}!}}\psi_{0}({\boldsymbol{x}},t)
    H_{{\boldsymbol{n}}}^{\left\{-\Lambda_{p}^{*}
    \Lambda_{p}^{-1}\right\}}\left(-\frac{i}{\hbar}(\Lambda_{p}^{+})^{-1}{\boldsymbol{x}}+
    \boldsymbol{\delta}-\Lambda_p{\left(\Lambda^*_p\right)}^{-1}\boldsymbol{\delta}^*\right),
\end{equation}
where $\psi_{0}({\boldsymbol{x}},t)$ is
$\psi_{{\boldsymbol{\alpha}}}({\boldsymbol{x}},t)$ with
${\boldsymbol{\alpha}}=0$.

We have shown that using time-dependent invariants method one can obtain the coherent and Fock
states of the system with quadratic Hamiltonian. The problem of finding these states reduces
to solving equations \eqref{e:LD} for the matrix $\Omega(t)$ and the vector
$\boldsymbol{\delta}(t)$. As an example we consider the class of systems for which $B_{px} =
B_{xp}=0$ and $B_{pp}$ and $B_{xx}$ are time-independent, permutable and non-singular. In this
case for matrices $\Lambda_p$ and $\Lambda_x$ we get evolution equations with constant
coefficients
\begin{align}
    \dot{\Lambda}_p &= -\Lambda_xB_{pp} + \Lambda_pB_{xp}, \\
    \dot{\Lambda}_x &= -\Lambda_xB_{px} + \Lambda_pB_{xx}, \\
\end{align}
or the system of equations in the form
\begin{equation}
    \begin{Vmatrix}\dot{\Lambda}^T_p \\ \\ \dot{\Lambda}^T_x\end{Vmatrix} =
    \begin{Vmatrix}B^T_{xp} & -B^T_{pp} \\ \\ B^T_{xx} & -B^T_{px}\end{Vmatrix}
    \begin{Vmatrix}\Lambda^T_p \\ \\ \Lambda^T_x\end{Vmatrix}.
\end{equation}
Taking into account the initial conditions
\begin{equation}
    \Lambda_p(0) = A_p, \qquad \Lambda_x(0) = A_x
\end{equation}
the solution of this system of equations reads
\begin{equation}\label{e:o}
    \begin{Vmatrix}\Lambda^T_p(t) \\ \\ \Lambda^T_x(t)\end{Vmatrix} =
    \exp\left\{\begin{Vmatrix}0 & -B_{pp} \\ \\ B_{xx} & 0\end{Vmatrix}t\right\}
    \begin{Vmatrix}A^T_p \\ \\ A^T_x\end{Vmatrix}.
\end{equation}
As shown in the Appendix B, the matrix exponent in right hand side of the equation \eqref{e:o}
reads
\begin{equation}
    \exp\left\{\begin{Vmatrix}0 & -B_{pp} \\ \\ B_{xx} &
    0\end{Vmatrix}t\right\} = \begin{Vmatrix}\cos\left(\sqrt{B_{pp}B_{xx}}\;t\right) & -\sqrt{B_{pp}B^{-1}_{xx}}\sin\left(\sqrt{B_{xx}B_{pp}}\;t\right) \\ \\
                   \sqrt{B_{xx}B^{-1}_{pp}}\sin\left(\sqrt{B_{pp}B_{xx}}\;t\right) &
                   \cos\left(\sqrt{B_{xx}B_{pp}}\;t\right)\end{Vmatrix}.
\end{equation}
Thus one get the explicit form of the solution \eqref{e:o} for the matrices
\begin{align}
    \Lambda_p(t) &= A_p\cos\left(\sqrt{B_{pp}B_{xx}}\;t\right) - A_x\sqrt{B_{pp}B^{-1}_{xx}}\sin\left(\sqrt{B_{xx}B_{pp}}\;t\right), \\
    \Lambda_x(t) &= A_p\sqrt{B_{xx}B^{-1}_{pp}}\sin\left(\sqrt{B_{pp}B_{xx}}\;t\right) +
    A_x\cos\left(\sqrt{B_{xx}B_{pp}}\;t\right).
\end{align}
in the partial case of the quadratic multimode systems of the considered example.

In this section we have shown that the wave function \eqref{e:coh} of the multimode coherent
state of the quantum systems with quadratic Hamiltonian is given by Gaussian which depends on
the parameters of symplectic transform which determine the linear invariants. We also have
shown that the wave function \eqref{e:psin} of Fick state of the multimode quadratic system is
expressed in terms of multivariale Hermite polynomials which depend on the parameters of the
symplectic transform, These two expressions for the wave functions of the coherent and Fock
states are the main results of the section.

\section{Coherent and Fock States in the Probability Representation}
In this section we discuss the new probability or tomographic representation of quantum
mechanics. We obtain the explicit expressions for tomograms of the coherent and Fock states of
the quadratic multimode systems discussed in previous section in framework of standard
approach. In \cite{qso-7-615} an operator $\widehat{\boldsymbol{X}} =
(\widehat{X}_{1},....,\widehat{X}_{N})$, where $n=1,...,N$, is discussed for N = 1 as a
generic linear combination of position and momentum operators
\begin{equation}\label{E:one}
    \hat{X}_{n} = \mu_{n}\hat{x}_{n} + \nu_{n}\hat{p}_{n},
\end{equation}
where $\mu_{n}$ and $\nu_{n}$ are real parameters for all n, and $\hat{{\boldsymbol{X}}}$ is
Hermetian, hence observable. The physical meaning of the vectors ${\boldsymbol{\mu}} =
(\mu_{1},...,\mu_{N})$ and ${\boldsymbol{\nu}} = (\nu_{1},...,\nu_{N})$ is that they describe
an ensemble of rotated and scaled reference frames, in classical phase space, in which the
position ${\boldsymbol{X}}$ may be measured. In \cite{qso-7-615} it is shown that the quantum
state of a system is completely determined if the classical probability distribution
$w({\boldsymbol{X}},{\boldsymbol{\mu}},{\boldsymbol{\nu}})$, for the variable
${\boldsymbol{X}}$ is given in an ensemble of reference frames in the classical phase space
(MDF). Such a function, also known as the marginal distribution function, belongs to a broad
class of distributions which are determined as the Fourier transform of a characteristic
function \cite{pr-177-1882}. For the particular case of the variable \eqref{E:one}, considered
in \cite{qso-7-615}, \cite{fp-27-801}, \cite{pl-a-213-1}, the scheme of \cite{pr-177-1882}
gives
\begin{equation}\label{E:two}
    w({\boldsymbol{X}},{\boldsymbol{\mu}},{\boldsymbol{\nu}})=\frac{1}{(2\pi\hbar)^{N}}
    \int\exp(-i{\boldsymbol{k}}^T{\boldsymbol{X}})\left\langle\exp(i{\boldsymbol{k}}^T\widehat{{\boldsymbol{X}}})\right\rangle\,d{\boldsymbol{k}},
\end{equation}
where $\langle A\rangle=Tr(\hat{\rho}\hat{A})$, $\hat{\rho}$ is the density operator. In
\cite{pr-177-1882} it was shown that, whenever $\hat{{\boldsymbol{X}}}$ is observable,
$w({\boldsymbol{X}},{\boldsymbol{\mu}},{\boldsymbol{\nu}})$ is indeed a probability
distribution. It is positive definite and satisfies the normalization condition
\begin{equation}\label{E:three}
    \int w({\boldsymbol{X}},{\boldsymbol{\mu}},{\boldsymbol{\nu}})\,d{\boldsymbol{X}} = 1.
\end{equation}
The definition of the MDF allows us to express it in terms of the
density matrix
\begin{align}\label{E:four}
\begin{split}
    w({\boldsymbol{X}},{\boldsymbol{\mu}},{\boldsymbol{\nu}}) &= \frac{1}{(2\pi
    \hbar )^{N}\vert\nu_{1}\cdot\ldots\cdot\nu_{N}\vert}\int
    \rho({\boldsymbol{z}},{\boldsymbol{z^{\prime}}})\prod_{n=1}^{N}\exp
    \left[ -i \frac{z_{n}-z^{\prime}_{n}}{\hbar\nu_{n}}
    \left( X_{n}-\mu_{n}\frac{z_{n}+z^{\prime}_{n}}{2}
    \right)\right]\,d{\boldsymbol{z}}\,d{\boldsymbol{z^{\prime}}}.
\end{split}
\end{align}
Recalling the relation between the Wigner function and the density matrix, the equation
\eqref{E:four} may be rewritten as a relation between
$w({\boldsymbol{X}},{\boldsymbol{\mu}},{\boldsymbol{\nu}})$ and the Wigner function,
\begin{equation}\label{E:five}
    w({\boldsymbol{X}},{\boldsymbol{\mu}},{\boldsymbol{\nu}}) =
    \int\prod_{n=1}^{N}\exp[-ik_{n}(X_{n}-\mu_{n}x_{n}-\nu_{n}p_{n})]W({\boldsymbol{x}},{\boldsymbol{p}})\,\frac{d{\boldsymbol{k}}\,d{\boldsymbol{p}}\,d{\boldsymbol{x}}}{(2\pi\hbar)^{N}}.
\end{equation}
Although the general class of distribution functions of the kind \eqref{E:two} was introduced,
as a function of the density matrix, already by Cahill and Glauber in \cite{pr-177-1882}, they
did not analyze the possibility of a new approach to quantum mechanics, in terms of such
distribution functions, mainly because the invertibility of \eqref{E:four} was not
investigated. An important step in this direction is represented by \cite{pr-a-40-2847} where
Vogel and Risken have shown that for a particular choice of the parameters $\mu_{n}$ and
$\nu_{n}$ the marginal distribution completely determines the Wigner function. We make our
previous statements more precise, saying that the MDF (the classical probability associated to
the random variable ${\boldsymbol{X}}$) contains the same information on a quantum system as
the density matrix. Marginal distribution
$w({\boldsymbol{X}},{\boldsymbol{\mu}},{\boldsymbol{\nu}})$ can be also expressed through wave
function of the system. This expression is given by
\begin{align}\label{e:wpsi}
\begin{split}
    w({\boldsymbol{X}},{\boldsymbol{\mu}},{\boldsymbol{\nu}},t) &= \frac{1}{(2\pi\hbar)^{N}
    \vert\nu_{1}\cdot\ldots\cdot\nu_{N}\vert}\left| \int\
    \psi({\boldsymbol{y}},t)\exp\biggl\{\frac{i}{2\hbar}{\boldsymbol{y}}^TMN^{-1}{\boldsymbol{y}}-\frac{i}{\hbar}{\boldsymbol{y}}^TN^{-1}{\boldsymbol{X}}\biggr\}
    \,d{\boldsymbol{y}} \right|^{2},
\end{split}
\end{align}
where the diagonal matrices which contain the parameters determining the reference frames in
the phase space of the multimode quadratic system
\begin{align}
    M &= \diag\{\mu_1,\ldots,\mu_N\}, \\
    N &= \diag\{\nu_1,\ldots,\nu_N\}.
\end{align}
are introduced. Looking at the generic tomogram \eqref{e:wpsi} it is obvious that to evaluate
the tomograms corresponding to coherent state
$\psi_{{\boldsymbol{\alpha}}}({\boldsymbol{x}},t)$ and the Fock state
$\psi_{{\boldsymbol{n}}}({\boldsymbol{x}},t)$ we need to evaluate the following two integrals:
\begin{equation}\label{e:71}
    I_{{\boldsymbol{\alpha}}}(t) = \int\psi_{{\boldsymbol{\alpha}}}({\boldsymbol{y}},t)
    \exp\biggl\{\frac{i}{2\hbar}{\boldsymbol{y}}^TMN^{-1}{\boldsymbol{y}}-
    \frac{i}{\hbar}{\boldsymbol{y}}^TN^{-1}{\boldsymbol{X}}\biggr\}\,d{\boldsymbol{y}}
\end{equation}
and
\begin{equation}\label{e:72}
    I_{{\boldsymbol{n}}}(t) = \int\psi_{{\boldsymbol{n}}}({\boldsymbol{y}},t)
    \exp\biggl\{\frac{i}{2\hbar}{\boldsymbol{y}}^TMN^{-1}{\boldsymbol{y}}-
    \frac{i}{\hbar}{\boldsymbol{y}}^TN^{-1}{\boldsymbol{X}}\biggr\}\,d{\boldsymbol{y}}.
\end{equation}
Remembering the expression for the generating function \eqref{e:psialph} we have the relation
\begin{equation}\label{e:Ia}
    I_{{\boldsymbol{\alpha}}}(t) = e^{-\frac{1}{2}\parallel
    {\boldsymbol{\alpha}}\parallel^2}\sum\limits_{{\boldsymbol{n}}=
    0}^{\infty}I_{{\boldsymbol{n}}}(t)\frac{{{\boldsymbol{\alpha}}}^{{\boldsymbol{n}}}}{\sqrt{{\boldsymbol{n}}!}}.
\end{equation}
which shows that the integral \eqref{e:71} is the generating function for the integrals
\eqref{e:72}. In view of \eqref{e:Ia} we can obtain expression for $I_{{\boldsymbol{n}}}$ in
terms of derivatives of $I_{{\boldsymbol{\alpha}}}$
\begin{equation}\label{e:I1}
    I_{{\boldsymbol{n}}}(t) =
    \frac{1}{\sqrt{{\boldsymbol{n}}!}}\left.\frac{\partial}{\partial{\boldsymbol{\alpha}}^{\boldsymbol{n}}}\right|_{\boldsymbol{\alpha}=0}
    \Bigl(e^{\frac{1}{2}\Vert{\boldsymbol{\alpha}}\Vert}I_{{\boldsymbol{\alpha}}}(t)\Bigr).
\end{equation}
where
\begin{equation}
    \frac{\partial}{\partial{\boldsymbol{\alpha}}^{\boldsymbol{n}}}=
    \frac{\partial}{\partial\alpha^{n_1}_1}\cdot\ldots\cdot\frac{\partial}{\partial\alpha^{n_N}_N}
\end{equation}
According to \eqref{e:coh} the wave function of multimode coherent state
$\psi_{{\boldsymbol{\alpha}}}({\boldsymbol{x}},t)$ can be represented as
\begin{equation}
    \psi_{{\boldsymbol{\alpha}}}({\boldsymbol{x}},t) = \Psi(t)\exp\biggl\{
    -\frac{1}{2\hbar}{\boldsymbol{x}}^T\Lambda_0{\boldsymbol{x}}+
    \frac{i}{\hbar}{\boldsymbol{x}}^T\Lambda^{-1}_p({\boldsymbol{\alpha}}-{\boldsymbol{\delta}})+
    K({\boldsymbol{\alpha}},t)\biggr\},
\end{equation}
where we introduced the matrix
\begin{equation}
    \Lambda_0 = i\Lambda^{-1}_p\Lambda_x
\end{equation}
and the function
\begin{align}
     K({\boldsymbol{\alpha}},t) &=
    \frac{1}{2}{\boldsymbol{\alpha}}^T\Lambda^*_p\Lambda^{-1}_p{\boldsymbol{\alpha}}+
    {\boldsymbol{\alpha}}^T\left({\boldsymbol{\delta}}^*-\Lambda^*_p\Lambda^{-1}_p{\boldsymbol{\delta}}\right)+
    \frac{1}{2}{\boldsymbol{\delta}}^T\Lambda^*_p\Lambda^{-1}_p{\boldsymbol{\delta}}\notag\\
    &-\frac{1}{2}{\Vert{\boldsymbol{\delta}}\Vert}^2+
    i\int\limits_{0}^{t} \ \im\left({\dot{{\boldsymbol{\delta}}}}^T{\boldsymbol{\delta}}^*\right)\,d\tau-
    \frac{1}{2}{\Vert{\boldsymbol{\alpha}}\Vert}^2.
\end{align}
which does not contain the dependence on $\boldsymbol{X}$-variable. Using the known formula
for multidimensional Gaussian integral
\begin{equation}\label{e:exp}
    \int\exp\biggl\{-\frac{1}{2}{\boldsymbol{x}}^TA{\boldsymbol{x}}+{\boldsymbol{x}}^T{\boldsymbol{a}}\biggr\}\,d{\boldsymbol{x}} =
    \sqrt{\frac{{(2\pi)}^N}{\det
    A}}\exp\biggl\{\frac{1}{2}{\boldsymbol{a}}^TA^{-1}{\boldsymbol{a}}\biggr\},
\end{equation}
we can evaluate the integral \eqref{e:71}. After some transformations using properties of the
matrices $\Lambda_p$ and $\Lambda_x$ we can obtain the following result:
\begin{equation}
   I_{{\boldsymbol{\alpha}}}(t)=\Psi(t)\sqrt{\frac{{(2\pi\hbar)}^N}{\det\left(\Lambda_0-iMN^{-1}\right)}}
    E({\boldsymbol{X}},{\boldsymbol{\mu}},{\boldsymbol{\nu}},{\boldsymbol{\alpha}},t),
\end{equation}
where we introduce the function
\begin{equation}
    E({\boldsymbol{X}},{\boldsymbol{\mu}},{\boldsymbol{\nu}},{\boldsymbol{\alpha}},t) =
    \exp\biggl\{-\frac{1}{2}Q({\boldsymbol{X}},{\boldsymbol{\mu}},{\boldsymbol{\nu}},{\boldsymbol{\alpha}},t)+K({\boldsymbol{\alpha}},t)\biggr\},
\end{equation}
which is determined by the function
\begin{equation}
    Q({\boldsymbol{X}},{\boldsymbol{\mu}},{\boldsymbol{\nu}},{\boldsymbol{\alpha}},t) = -\frac{1}{\hbar}{\left({\boldsymbol{X}}-N\Lambda^{-1}_p({\boldsymbol{\alpha}}-{\boldsymbol{\delta}})\right)}^T
    N^{-1}{\left(\Lambda_0-iMN^{-1}\right)}^{-1}N^{-1}\left({\boldsymbol{X}}-N\Lambda^{-1}_p({\boldsymbol{\alpha}}-{\boldsymbol{\delta}})\right).
\end{equation}
For the tomogram of the multimode coherent state
$w_{{\boldsymbol{\alpha}}}({\boldsymbol{X}},{\boldsymbol{\mu}},{\boldsymbol{\nu}},t)$, we get
the expression
\begin{equation}
    w_{{\boldsymbol{\alpha}}}({\boldsymbol{X}},{\boldsymbol{\mu}},{\boldsymbol{\nu}},t) =
    \frac{1}{{(2\pi\hbar^2)}^{N/2}\vert\nu_1\cdot\ldots\cdot\nu_N\vert}
     \frac{\vert\det N\vert}{\sqrt{\det\Xi^+({\boldsymbol{\mu}},{\boldsymbol{\nu}},t)\Xi({\boldsymbol{\mu}},{\boldsymbol{\nu}},t)}}{\vert
     E({\boldsymbol{X}},{\boldsymbol{\mu}},{\boldsymbol{\nu}},{\boldsymbol{\alpha}},t)\vert}^2,
\end{equation}
where the matrix $\Xi({\boldsymbol{\mu}},{\boldsymbol{\nu}},t)$ is given by
\begin{equation}
    \Xi({\boldsymbol{\mu}},{\boldsymbol{\nu}},t) = i\Lambda_x(t)N-i\Lambda_p(t)M.
\end{equation}
On the other hand, we can represent
$w_{{\boldsymbol{\alpha}}}({\boldsymbol{X}},{\boldsymbol{\mu}},{\boldsymbol{\nu}},t)$ as
multidimensional Gaussian distribution function, i.e
\begin{equation}\label{e:H}
    w_{{\boldsymbol{\alpha}}}({\boldsymbol{X}},{\boldsymbol{\mu}},{\boldsymbol{\nu}},t) =
    \frac{1}{\sqrt{{(2\pi)}^N\det\Sigma}}\exp\biggl\{
    -\frac{1}{2}{\left({\boldsymbol{X}}-{\boldsymbol{X}}_0\right)}^T
    \Sigma^{-1}\left({\boldsymbol{X}}-{\boldsymbol{X}}_0\right)\biggr\},
\end{equation}
where the dispertion matrix depends on the reference frame parameters and the parameters of
symplectic transform
\begin{equation}
    \Sigma({\boldsymbol{\mu}},{\boldsymbol{\nu}},t) = \hbar^2\Xi^+({\boldsymbol{\mu}},{\boldsymbol{\nu}},t)\Xi({\boldsymbol{\mu}},{\boldsymbol{\nu}},t),
\end{equation}
and the mean value of the multidimensional random variable
\begin{equation}
    {\boldsymbol{X}}_0({\boldsymbol{\mu}},{\boldsymbol{\nu}},{\boldsymbol{\alpha}},t) = \hbar\Xi^+({\boldsymbol{\mu}},{\boldsymbol{\nu}},t)\Xi({\boldsymbol{\mu}},{\boldsymbol{\nu}},t)\left(\Xi^{-1}({\boldsymbol{\mu}},{\boldsymbol{\nu}},t)({\boldsymbol{\alpha}}
    -{\boldsymbol{\delta}}(t))+(\Xi^*)^{-1}({\boldsymbol{\mu}},{\boldsymbol{\nu}},t)({\boldsymbol{\alpha}}^*-{\boldsymbol{\delta}}^*(t))\right).
\end{equation}
is determined also be the symplectic transform parameters and depends on the reference frame
parameters. Using expression for the tomogram
$w_{{\boldsymbol{\alpha}}}({\boldsymbol{X}},{\boldsymbol{\mu}},{\boldsymbol{\nu}},t)$ and
formula for the integral \eqref{e:I1} we can obtain tomograms of the Fock state:
\begin{align} \label{e:wn}
\begin{split}
    w_{{\boldsymbol{n}}}({\boldsymbol{X}},{\boldsymbol{\mu}},{\boldsymbol{\nu}},t) &=
    w_{{\boldsymbol{0}}}({\boldsymbol{X}},{\boldsymbol{\mu}},{\boldsymbol{\nu}},t)\frac{1}{{\boldsymbol{n}}!}
    {\left\vert H^{\left\{(\Xi^T)^{-1}\Xi^+\right\}}_{{\boldsymbol{n}}}\left(\frac{1}{\hbar}(\Xi^+)^{-1}{\boldsymbol{X}}+
    (\Xi^+)^{-1}\left(\Xi^+{\boldsymbol{\delta}}+\Xi^T{\boldsymbol{\delta}}^*\right)\right)\right\vert}^2,
\end{split}
\end{align}
where $w_{{\boldsymbol{0}}}({\boldsymbol{X}},{\boldsymbol{\mu}},{\boldsymbol{\nu}},t)$ is $
w_{{\boldsymbol{\alpha}}}({\boldsymbol{X}},{\boldsymbol{\mu}},{\boldsymbol{\nu}},t)$ with
${\boldsymbol{\alpha}} = 0$. For one-dimensional case this formula takes the form
\begin{equation}\label{e:walph1}
    w_n(X,\mu,\nu,t) = w_0(X,\mu,\nu,t)\frac{1}{2^nn!}
    {\left\vert H_n\left(\frac{1}{\hbar}\frac{X}{\sqrt{2}\vert\xi\vert}+
    \frac{\xi^*\delta+\xi\delta^*}{\sqrt{2}\vert\xi\vert}\right)\right\vert}^2,
\end{equation}
where the matrix $\Xi$, is the one-dimensional matrix, and it is denoted as $\xi$. The numbers
$\xi$ and $\delta$ depend on time.

The explicit expression of the tomogram of the Fock state of the multimode quadratic systems
in terms of the multivariable Hermite polynomials \eqref{e:wn} is the main result of this
section.

\section{New relations for Hermite Polynomials}
In this Section we get some new formulas for multivariable Hermite polynomials using results
obtained in previous Sections. In section 1, considering the problem of finding coherent and
Fock states, we have found these states for matrices $A_p$ and $A_x$ which satisfy conditions
\eqref{e:A1} and \eqref{e:nA2}. If we chose other matrices $A'_p$ and $A'_x$ which also
satisfy the same conditions, we obtain other coherent states and other Fock states. We want to
find out how "new" Fock state $\psi^{\prime}_{{\boldsymbol{n}}}$ with matrices $A'_p$ and
$A'_x$ can be expressed through "old" Fock states $\psi_{{\boldsymbol{m}}}$ with matrices
$A_p$ and $A_x$. We consider the case $\boldsymbol{c} = 0$ (hence $\boldsymbol{\delta} = 0$).

Making symplectic transformation of annihilation and creation operators ${\boldsymbol{a}}$ and
${\boldsymbol{a}}^{+}$ of the form
\begin{equation}\label{e:trans}
    \begin{pmatrix}\widehat{{\boldsymbol{a}}'} \\ {\widehat{{\boldsymbol{a}}'}}^+\end{pmatrix} =
     S
    \begin{pmatrix}\widehat{{\boldsymbol{a}}} \\
    \widehat{{\boldsymbol{a}}}^+\end{pmatrix},\qquad
    S = \begin{Vmatrix}S_p & S_x \\ \\ S^*_x & S^*_p\end{Vmatrix},
\end{equation}
we obtain "new" operators $\widehat{{\boldsymbol{a}}'}$ and
${\widehat{{\boldsymbol{a}}'}}^{+}$. These operators must satisfy condition \eqref{e:a2} and
therefore the matrix $S$ must satisfy the following condition:
\begin{equation}
    S\Sigma_{2N}S^T = \Sigma_{2N},
\end{equation}
or
\begin{align}
    S_pS^T_x-S_xS^T_p &= S^*_pS^+_x-S^*_xS^+_p = 0, \label{e:S1}\\
    S_pS^+_p-S_xS^+_x &= S^*_pS^T_p-S^*_xS^T_x = E_N \label{e:S2}.
\end{align}
It is easy to show that $N \times N$-matrix $S_p$ is non-singular. In fact, let
${\boldsymbol{z}}\not=0$ is the vector for which $S^+_p{\boldsymbol{z}} = 0$. It means that
the matrix $S_p$ has no inverse, i.e. it is singular one. In this case equation \eqref{e:S2}
gives for scalar product of vectors the inequality
\begin{equation}
    \left({\boldsymbol{z}}, S_pS^+_p{\boldsymbol{z}}\right) - \left({\boldsymbol{z}}, S_xS^+_x{\boldsymbol{z}}\right) =
    -\left(S^+_x{\boldsymbol{z}}, S^+_x{\boldsymbol{z}}\right) = -{\Vert S^+_x{\boldsymbol{z}}\Vert}^2 = {\Vert{\boldsymbol{z}}\Vert}^2
    > 0.
\end{equation}
This inequality contradicts to the inequality $-{\Vert S^+_x{\boldsymbol{z}}\Vert}^2 \leqslant
0 $. Thus, we proved the nonsingularity of the matrix $S_p$.

The functions $\psi_{{\boldsymbol{n}}}({\boldsymbol{x}},t)$ form complete set of function and
we can represent the function $\psi'_{{\boldsymbol{n}}}({\boldsymbol{x}},t)$ as the series
\begin{equation}
    \psi^{\prime}_{{\boldsymbol{n}}}({\boldsymbol{x}},t) =
    \sum^{+\infty}\limits_{{\boldsymbol{m}}=0}c_{{\boldsymbol{n}}{\boldsymbol{m}}}\psi_{{\boldsymbol{m}}}({\boldsymbol{x}},t).
\end{equation}
The complex coefficients $c_{\boldsymbol{n}\boldsymbol{m}}$ have the physical meaning of
transition amplitude from an initial Fock state $|\boldsymbol{n}\rangle^{\prime}$  to a final
Fock state  $|\boldsymbol{m}\rangle$. According to \eqref{e:psin} and \eqref{e:trans} "new"
Fock state $ \psi'_{{\boldsymbol{n}}}({\boldsymbol{x}},t)$ is given explicitly by the formula
\begin{align}
\begin{split}
    \psi^{\prime}_{{\boldsymbol{n}}}({\boldsymbol{x}},t) &= {\left(2\pi\hbar^2\right)}^{-N/4}
    {\left(\det\left(S_pA_p+S_xA^*_p\right)\right)}^{-1/2}
    \exp\left\{-\frac{i}{2\hbar}{\boldsymbol{x}}^T{\left(S_pA_p+S_xA^*_p\right)}^{-1}
    \left(S_pA_x+S_xA^*_x\right){\boldsymbol{x}}\right\}\\
    &\times\frac{1}{\sqrt{{\boldsymbol{n}}!}}
    H^{\left\{-\left(S^*_pA^*_p+S^*_xA_p\right){\left(S_pA_p+S_xA^*_p\right)}^{-1}\right\}}_{{\boldsymbol{n}}}
    \left(-\frac{i}{\hbar}{\left(A^+_pS^+_p+A^T_pS^+_x\right)}^{-1}{\boldsymbol{x}}\right).
\end{split}
\end{align}
As it is known the decomposition coefficients $c_{{\boldsymbol{n}}{\boldsymbol{m}}}$ are given
by scalar product of wave functions $\psi_{{\boldsymbol{m}}}({\boldsymbol{x}},t)$ and
$\psi^{\prime}_{{\boldsymbol{n}}}({\boldsymbol{x}},t)$ which is expressed in terms of the
overlap integral of two multivariable Hermite polynomials and Gaussian of the form
\begin{align}
\begin{split}
    c_{{\boldsymbol{n}}{\boldsymbol{m}}} &= \bigl\langle\psi_{{\boldsymbol{m}}}\bigm|\psi^{\prime}_{{\boldsymbol{n}}}\bigr\rangle =
    {\left(2\pi\hbar^2\right)}^{-N/2}{\left(\det\left(S_pA_p+S_xA^*_p\right)\right)}^{-1/2}
    {\left(\det A^*_p\right)}^{-1/2}\frac{1}{\sqrt{{\boldsymbol{n}}!}}\frac{1}{\sqrt{{\boldsymbol{m}}!}}\\
    &\times\int\exp\Biggl\{-\frac{i}{2\hbar}{\boldsymbol{x}}^T\Bigl(
    -{\left(A^*_p\right)}^{-1}A^*_x+{\left(S_pA_p+S_xA^*_p\right)}^{-1}
    \left(S_pA_x+S_xA^*_x\right)\Bigr){\boldsymbol{x}}\Biggr\}\\
    &\times H^{\left\{-\left(S^*_pA^*_p+S^*_xA_p\right){\left(S_pA_p+S_xA^*_p\right)}^{-1}\right\}}_{{\boldsymbol{n}}}
    \left(-\frac{i}{\hbar}{\left(A^+_pS^+_p+A^T_pS^+_x\right)}^{-1}{\boldsymbol{x}}\right)
    H^{\left\{-A_p{\left(A^*_p\right)}^{-1}\right\}}_{{\boldsymbol{m}}}\left(-\frac{i}{\hbar}{\left(A^T_p\right)}^{-1}{\boldsymbol{x}}\right)\,d{\boldsymbol{x}}.
\end{split}
\end{align}
Here we used Dirac notations
\begin{equation}
    \bigl\langle \varphi \bigm| \psi \bigr\rangle  =
    \int \varphi^*({\boldsymbol{x}}) \psi({\boldsymbol{x}})\,d{\boldsymbol{x}}
\end{equation}
for the standard scalar product of two vectors $\bigl|\varphi\bigr\rangle,
\bigl|\psi\bigr\rangle \in L_2(\mathbf{C}^N)$ written in position representation. Evaluating
this integral we obtain for the transition amplitude
\begin{align}\label{e:coff}
\begin{split}
    c_{{\boldsymbol{n}}{\boldsymbol{m}}} &= \frac{\sqrt{{\boldsymbol{n}}!{\boldsymbol{m}}!}}{\sqrt{\det S_p}}
    \left[{\boldsymbol{t}}^{{\boldsymbol{n}}}{\boldsymbol{s}}^{{\boldsymbol{m}}}\right]
    \exp\Biggl\{\frac{1}{2}{\boldsymbol{t}}^TS^*_xS^{-1}_p{\boldsymbol{t}}+{\boldsymbol{s}}^TS^{-1}_p{\boldsymbol{t}}-
    \frac{1}{2}{\boldsymbol{s}}^TS^{-1}_pS_x{\boldsymbol{s}}\Biggr\}
    = \frac{1}{\sqrt{\det
    S_p}\sqrt{{\boldsymbol{n}}!{\boldsymbol{m}}!}}H^{\{F\}}_{({\boldsymbol{n}},{\boldsymbol{m}})}(0,0),
\end{split}
\end{align}
where the $2N \times 2N$-matrix
\begin{equation}
F = \begin{Vmatrix}
    -S_x^*S^{-1}_p   &   -{\left(S^T_p\right)}^{-1} \\ \\
    -S^{-1}_p        &   S^{-1}_pS_x
    \end{Vmatrix}
\end{equation}
is expressed in terms of four block-matrices. The normalization condition of
$\psi'_{{\boldsymbol{n}}}({\boldsymbol{x}},t)$ is reduced to the equality
$\sum^{+\infty}\limits_{{\boldsymbol{m}}=0}{\left\vert
c_{{\boldsymbol{n}}{\boldsymbol{m}}}\right\vert}^2=1$. Using the expression \eqref{e:coff} for
coefficients $c_{{\boldsymbol{n}}{\boldsymbol{m}}}$ we can represent this equality in the
following from:
\begin{equation}\label{e:2N}
    \left\vert\det S_p\right\vert = \frac{1}{{\boldsymbol{n}}!}\sum\limits^{+\infty}_{{\boldsymbol{m}}=0}
    \frac{{\left\vert
    H^{\{F\}}_{({\boldsymbol{n}},{\boldsymbol{m}})}(0,0)\right\vert}^2}{{\boldsymbol{m}}!}.
\end{equation}
This form gives a sum rule for multivariable Hermite polynomials. Let us consider the partial
case ${\boldsymbol{n}} = 0$. From \eqref{e:coff} we obtain transition amplitude from ground
state $|0\rangle$ to an excited state $|\boldsymbol{m}\rangle$.
\begin{equation}
  c_{{\boldsymbol{0}}{\boldsymbol{m}}}=
  \frac{H^{\left\{S^{-1}_pS_x\right\}}_{{\boldsymbol{m}}}(0)}{\sqrt{\det
  S_p}\sqrt{{\boldsymbol{m}}!}}.
\end{equation}
The expansion
\begin{equation}
    \psi^{\prime}_0({\boldsymbol{x}},t) = \sum^{+\infty}\limits_{{\boldsymbol{m}}=
    0}c_{0{\boldsymbol{m}}}\psi_{{\boldsymbol{m}}}({\boldsymbol{x}},t)
\end{equation}
takes the explicit form
\begin{align}
\begin{split}
    &\exp\Biggl\{\frac{1}{2\hbar^2}{\boldsymbol{x}}^T{\left(S_pA_p+S_xA^*_x\right)}^{-1}S_x{\left(A^T_p\right)}^{-1}{\boldsymbol{x}}\Biggr\}=
    \sqrt{\det\left(E_N + S_xA^*_pA^{-1}_pS^{-1}_p\right)}\\
    &\times\sum\limits_{{\boldsymbol{m}}\geqslant 0}\frac{H^{\left\{S^{-1}_pS_x\right\}}_{{\boldsymbol{m}}}(0)}{{\boldsymbol{m}}!}
    H^{\left\{-A^*_pA^{-1}_p\right\}}_{{\boldsymbol{m}}}\left(-\frac{i}{\hbar}{\left(A^+_p\right)}^{-1}{\boldsymbol{x}}\right).
\end{split}
\end{align}
The obtained result provides the partial case of sum rule for multivariable Hermite
polynomials. Normalization condition for the new ground state of the quadratic system
$\psi^{\prime}_0({\boldsymbol{x}},t)$ is equivalent to the equality
\begin{equation}\label{e:detS}
    \left\vert\det S_p\right\vert = \sum^{+\infty}\limits_{{\boldsymbol{m}}=0}\frac{{\left\vert
    H^{\left\{S^{-1}_pS_x\right\}}_{{\boldsymbol{m}}}(0)\right\vert}^2}{{\boldsymbol{m}}!}
\end{equation}
which is new sum rule for the Hermite polynomials. We can verify that this equality is true in
some particular case. For example, if $N=1$ and the matrix
\begin{equation} \label{e:mS}
    S = \begin{Vmatrix}\cosh\theta & \sinh\theta \\
                       \sinh\theta & \cosh\theta \end{Vmatrix},
\end{equation}
i.e.,  $S_p = \cosh\theta$, $S_x = \sinh\theta$, the Hermite polynomials reads
\begin{equation}
    H^{\{\tanh\theta\}}_n(0) =
    {\left(\frac{\tanh\theta}{2}\right)}^{n/2}H_n(0) =
    \begin{cases}
        0 & n = 2m+1 \\
        {(-1)}^m\frac{{(2m)}!}{2^mm!}{\left(\tanh\theta\right)}^m & n = 2m
    \end{cases}
\end{equation}
and we obtain from \eqref{e:detS} the decomposition of the form
\begin{equation} \label{e:expan}
    \cosh\theta =
    \sum\limits^{+\infty}_{m=0}\frac{(2m)!}{2^{2m}{(m!)}^2}{\left(\tanh\theta\right)}^{2m}.
\end{equation}
Taking into account that $\cosh\theta = 1/\sqrt{1-{\left(\tanh\theta\right)}^2}$ we can
conclude that \eqref{e:expan} is true, since expansion of function $1/\sqrt{1-x}$ is given by
the obvious relation
\begin{equation}
    \frac{1}{\sqrt{1-x}} = \sum\limits^{+\infty}_{m=0}
    \frac{(2m)!}{2^{2m}{(m!)}^2}x^m.
\end{equation}
Let us consider the case $n\not=0$.  Using the relation between two-dimensional Hermite
polynomials and Legendre polynomials:
\begin{align}
\begin{split}
    H^{\{R\}}_{nm}(0,0) =
    (\min(n,m))!{(-1)}^{(n+m)/2}r^{n/2}_{11}r^{m/2}_{22}
    {\left(\frac{r^2_{12}}{r_{11}r_{22}}-1\right)}^{(n+m)/4}
    P^{\vert n-m\vert/2}_{(n+m)/2}\left(\frac{r_{12}}{r^2_{12}-r_{11}r_{22}}\right),
\end{split}
\end{align}
where the $2 \times 2$-matrix $R$ has the form
\begin{equation}\label{e:R}
    R =
    \begin{Vmatrix}
         r_{11}           & r_{12}   \\
          r_{21}           &  r_{22}
    \end{Vmatrix},
\end{equation}
we obtain for matrix $S$ given by \eqref{e:mS} the expression
\begin{equation}
    H^{\{R\}}_{({\boldsymbol{n}},{\boldsymbol{m}})}(0,0) =
       (\min(n,m))!{(-1)}^{5n/4+3m/2}P^{\vert n-m\vert/2}_{(n+m)/2}
       \left(\frac{1}{\cosh\theta}\right),
\end{equation}
where the matrix elements in \eqref{e:R} are taken in the form
\begin{equation}
    R =
    \begin{Vmatrix}
        -\tanh\theta             &   -\frac{1}{\cosh\theta} \\ \\
        -\frac{1}{\cosh\theta}   &   \tanh\theta
    \end{Vmatrix}.
\end{equation}
Comparing the expression for transition probabilities obtained in probability representation
of quantum mechanics and in usual representation we can find other new relation for Hermite
polynomials. Transition probability in usual representation is given by the known relation
\begin{equation}
 w_{{\boldsymbol{n}}{\boldsymbol{m}}} =
 {\left\vert\bigl\langle\psi_{{\boldsymbol{n}}}\left(t_1\right)\bigm|
    \psi_{{\boldsymbol{m}}}\left(t_2\right)\bigr\rangle\right\vert}^2
\end{equation}
Therefore to represent the transition probability in explicit form we need to calculate the
following overlap integral of two wave functions:
\begin{align}
\begin{split}
    \bigl\langle\psi_{{\boldsymbol{n}}}\left(t_1\right)\bigm|
    \psi_{{\boldsymbol{m}}}\left(t_2\right)\bigr\rangle &=
    \int\psi_{{\boldsymbol{n}}}\left({\boldsymbol{x}},
    t_1\right)\psi_{{\boldsymbol{m}}}\left({\boldsymbol{x}},t_2\right)\,d{\boldsymbol{x}}
    =\frac{1}{\sqrt{{\left(i\hbar\right)}^N\det D}}
    \exp\biggl\{\frac{1}{2}{\boldsymbol{\delta}}^T\left(t_1\right)D^{-1}E{\boldsymbol{\delta}}\left(t_1\right)\biggr.\\
    \biggl.&+\frac{1}{2}{{\boldsymbol{\delta}}^*}^T\left(t_2\right)E^*D^{-1}{\boldsymbol{\delta}}^*\left(t_1\right)-
    \frac{i}{\hbar}{\boldsymbol{\delta}}^T\left(t_1\right)D^{-1}{\boldsymbol{\delta}}^*\left(t_2\right)\biggr.\\
    \biggl.&-\frac{1}{2}{\left\Vert{\boldsymbol{\delta}}\left(t_1\right)\right\Vert}^2-\frac{1}{2}{\left\Vert{\boldsymbol{\delta}}\left(t_2\right)\right\Vert}^2+
    i\int\limits^{t_1}_{t_2}\im\left(\dot{{\boldsymbol{\delta}}}^T{\boldsymbol{\delta}}^*\right)\,d\tau\biggr\}
    \frac{1}{\sqrt{{\boldsymbol{n}}!{\boldsymbol{m}}!}}H^{\{R\}}_{({\boldsymbol{n}},{\boldsymbol{m}})}({\boldsymbol{u}},{\boldsymbol{v}}),
\end{split}
\end{align}
where we introduced new $2N \times 2N$-matrix
\begin{equation}
    R = R(t_1,t_2) = -
    \begin{Vmatrix}
        D^{-1}E                                   &   -\displaystyle\frac{i}{\hbar}D^{-1} \\ \\
        -\displaystyle\frac{i}{\hbar}{\left(D^T\right)}^{-1}   &   E^*D^{-1}              \\
    \end{Vmatrix},
\end{equation}
and the $2N$-vector
\begin{equation}\label{e:127}
    \begin{pmatrix}
        {\boldsymbol{u}} \\
        {\boldsymbol{v}}
    \end{pmatrix}
    =
    \begin{pmatrix}
        {\boldsymbol{u}}(t_1,t_2) \\
        {\boldsymbol{v}}(t_1,t_2)
    \end{pmatrix}
    =
    \begin{Vmatrix}
        -E^+{\left(D^+\right)}^{-1}               &   -\displaystyle\frac{i}{\hbar}{\left(D^*\right)}^{-1} \\ \\
        -\displaystyle\frac{i}{\hbar}{\left(D^+\right)}^{-1}   &   -E{\left(D^*\right)}^{-1} \\
    \end{Vmatrix}
    \begin{pmatrix}
        D^{-1}E{\boldsymbol{\delta}}\left(t_1\right)-{\boldsymbol{\delta}}^*\left(t_1\right)-\displaystyle\frac{i}{\hbar}D^{-1}{\boldsymbol{\delta}}^*\left(t_2\right) \\ \\
        E^*D^{-1}{\boldsymbol{\delta}}^*\left(t_2\right)-{\boldsymbol{\delta}}\left(t_2\right)-\displaystyle\frac{i}{\hbar}D^{-1}{\boldsymbol{\delta}}\left(t_1\right)
    \end{pmatrix},
\end{equation}
The $N \times N$-matrices
\begin{align}
    D &= D(t_1,t_2) = \Lambda_p^*\left(t_2\right)\Lambda_x^T\left(t_1\right)-\Lambda_x^*\left(t_2\right)\Lambda_p^T\left(t_1\right), \\ \notag \\
    E &= E(t_1,t_2) = \Lambda_p^*\left(t_2\right)\Lambda_x^+\left(t_1\right)-\Lambda_x^*\left(t_2\right)\Lambda_p^+\left(t_1\right).
\end{align}
Are used to define the $2N$-vector \eqref{e:127}, and these matrices are connected with
symplectic matrix determining the linear integrals of motion of the system under
consideration. On the other hand, the transition probability from the initial state
$\bigl|{\boldsymbol{n}}\bigr\rangle$ to the final state $\bigl|{\boldsymbol{m}}\bigr\rangle$
in probability representation is given by the expression
\begin{equation} \label{e:trprob}
    w_{{\boldsymbol{n}}{\boldsymbol{m}}}
    =\frac{1}{{(2\pi)}^N}\int
    w_{{\boldsymbol{n}}}\left({\boldsymbol{X}}^{\prime},{\boldsymbol{\mu}},{\boldsymbol{\nu}},t_1\right)
    w_{{\boldsymbol{m}}}\left({\boldsymbol{X}}^{\prime\prime},-{\boldsymbol{\mu}},-{\boldsymbol{\nu}},t_2\right)
    e^{i\left({\boldsymbol{X}}^{\prime}+{\boldsymbol{X}}^{\prime\prime}\right)}
    \,d{\boldsymbol{\mu}}\,d{\boldsymbol{\nu}}\,d{\boldsymbol{X}}^{\prime}\,d{\boldsymbol{X}}^{\prime\prime},
\end{equation}
where we used the notation
\begin{equation}
    e^{{\boldsymbol{x}}} = \prod\limits^{N}_{k=1}e^{x_k} =
    e^{\sum\limits^N_{k=1}x_k}.
\end{equation}
Using expression \eqref{e:wn} obtained for the tomogram of the Fock state
$w_{n}({\boldsymbol{X}},{\boldsymbol{\mu}},{\boldsymbol{\nu}})$ and substituting it into
\eqref{e:trprob}, we obtain the equality
\begin{align}\label{e:H2}
\begin{split}
    &\int\left|H^{\left\{{\left(\Xi^T(t_1)\right)}^{-1}\Xi^+(t_1)\right\}}_{{\boldsymbol{n}}}\left(\frac{1}{\hbar}{\left(\Xi^+(t_1)\right)}^{-1}{\boldsymbol{X}}^{\prime}
    +{\left(\Xi^+(t_1)\right)}^{-1}\Bigl(\Xi^+(t_1){\boldsymbol{\delta}}(t_1)+\Xi^T(t_1){\boldsymbol{\delta}}^*(t_1)\Bigr)\right)\right|^2\\
    &\times\left|H^{\left\{{\left(\Xi^T(t_2)\right)}^{-1}\Xi^+(t_2)\right\}}_{{\boldsymbol{m}}}\left(-\frac{1}{\hbar}{\left(\Xi^+(t_2)\right)}^{-1}{\boldsymbol{X}}^{\prime\prime}
    +{\left(\Xi^+(t_2)\right)}^{-1}\Bigl(\Xi^+(t_2){\boldsymbol{\delta}}(t_2)+\Xi^T(t_2){\boldsymbol{\delta}}^*(t_2)\Bigr)\right)\right|^2\\
    &\times w_0\left({\boldsymbol{X}}^{\prime},{\boldsymbol{\mu}},{\boldsymbol{\nu}},t_1\right)w_0\left({\boldsymbol{X}}^{\prime\prime},-{\boldsymbol{\mu}},-{\boldsymbol{\nu}},t_2\right)
    e^{-i\left({\boldsymbol{X}}^{\prime}+{\boldsymbol{X}}^{\prime\prime}\right)}\,d{\boldsymbol{\mu}}\,d{\boldsymbol{\nu}}\,d{\boldsymbol{X}}^{\prime}\,d{\boldsymbol{X}}^{\prime\prime}
    =\frac{1}{\hbar^N\det D}\\
    &\times\left|\exp\biggl\{-\frac{1}{2}\left({\boldsymbol{\delta}}^T\left(t_1\right), {{\boldsymbol{\delta}}^*\left(t_2\right)}^T\right)R
    \begin{pmatrix}{\boldsymbol{\delta}}(t_1) \\
    {\boldsymbol{\delta}}^*(t_2)\end{pmatrix}
    -\frac{1}{2}{\left\Vert{\boldsymbol{\delta}}\left(t_1\right)\right\Vert}^2-\frac{1}{2}{\left\Vert{\boldsymbol{\delta}}\left(t_2\right)\right\Vert}^2\biggr\}
    \frac{1}{\sqrt{{\boldsymbol{n}}!{\boldsymbol{m}}!}}H^{\{R\}}_{({\boldsymbol{n}},{\boldsymbol{m}})}({\boldsymbol{u}},{\boldsymbol{v}})\right|^2.
\end{split}
\end{align}
For the partial case ${\boldsymbol{\delta}} = 0$, this general formula gives
\begin{align}
\begin{split}
    &\int{\left\vert H^{\left\{{\left(\Xi^T(t_1)\right)}^{-1}\Xi^+(t_1)\right\}}_{{\boldsymbol{n}}}\left(\frac{1}{\hbar}{\left(\Xi^+(t_1)\right)}^{-1}{\boldsymbol{X}}^{\prime}\right)\right\vert}^2
    {\left\vert H^{\left\{{\left(\Xi^T(t_2)\right)}^{-1}\Xi^+(t_2)\right\}}_{{\boldsymbol{m}}}\left(-\frac{1}{\hbar}{\left(\Xi^+(t_2)\right)}^{-1}{\boldsymbol{X}}^{\prime\prime}\right)\right\vert}^2\\
    &\times w_0\left({\boldsymbol{X}}^{\prime},{\boldsymbol{\mu}},{\boldsymbol{\nu}},t_1\right)w_0\left({\boldsymbol{X}}^{\prime\prime},-{\boldsymbol{\mu}},-{\boldsymbol{\nu}},t_2\right)
    e^{i\left({\boldsymbol{X}}^{\prime}+{\boldsymbol{X}}^{\prime\prime}\right)}\,d{\boldsymbol{\mu}}\,d{\boldsymbol{\nu}}\,d{\boldsymbol{X}}^{\prime}\,d{\boldsymbol{X}}^{\prime\prime}\\
    &=\frac{1}{\hbar^N\det D}
    \frac{1}{\sqrt{{\boldsymbol{n}}!{\boldsymbol{m}}!}}{\left\vert
    H^{\{R\}}_{({\boldsymbol{n}},{\boldsymbol{m}})}(0,0)\right\vert}^2.
\end{split}
\end{align}
The obtained two formulas are new mathematical results which are obvious from the point of
view of quantum transition consideration. But purely mathematical derivation of the obtained
formulas for multivariable Hermite polynomials is not that easy problem.

\section{Examples of Systems with Quadratic Hamiltonian}

\subsection{One-dimensional Harmonic Oscillator}
In this section, we consider some particular cases of systems with quadratic Hamiltonian, such
as harmonic oscillator, particle in electric field. We calculate tomograms of coherent and
Fock states of the system.

Our first example is a harmonic oscillator with time-dependent frequency and driving force.
This oscillator is described by the nonstationary Hamiltonian
\begin{equation}
    \widehat{H}(t) =
    \frac{\widehat{p}^2}{2m}+\frac{m{\omega(t)}^2\widehat{x}^2}{2}+f(t)\widehat{x}.
\end{equation}
This Hamiltonian is used to describe trapped ion. For the Hamiltonian the matrices $\Lambda_p$
used in previous sections and $\Lambda_x$ are given by the expressions
\begin{align}
    \Lambda_p &= \frac{i}{\sqrt{2m\omega\hbar}}\varepsilon, \\
    \Lambda_x &= -\frac{i}{\sqrt{2m\omega\hbar}}m\dot{\varepsilon},
\end{align}
where the function of time $\varepsilon$ satisfies the equation
\begin{equation}
    \ddot{\varepsilon} + \omega^2\varepsilon = 0.
\end{equation}
The initial conditions are taken in the form
\begin{equation}
    \varepsilon(0)=1,\qquad\dot{\varepsilon}(0)=\omega(0) i.
\end{equation}
For the one-dimensional matrix $\Xi$, dispersion matrix $\Sigma$ and vector $X_0$, we have
\begin{align}
    \Xi &= \frac{1}{\sqrt{2m\omega\hbar}}(m\dot{\varepsilon}\nu+\varepsilon\mu)\label{e:Xi}, \\
    \Sigma &= \frac{\hbar}{2m\omega}{\left\vert m\dot{\varepsilon}\nu+\varepsilon\mu\right\vert}^2 \label{e:Sigma},\\
    X_0 &= \sqrt{\frac{\hbar}{2m\omega}}\frac{1}{{\left\vert m\dot{\varepsilon}\nu+\varepsilon\mu\right\vert}^2}
           \Bigl((m\dot{\varepsilon}^*\nu+\varepsilon^*\mu)(\alpha-\delta)+
               (m\dot{\varepsilon}\nu+\varepsilon\mu)(\alpha^*-\delta^*)\Bigr)\label{e:X0}.
\end{align}
In this particular case all the matrices and vectors are one-dimensional objects. The wave
functions of coherent states and Fock states functions are given in explicit form
\begin{align}
    \psi_\alpha(x,t) &= \sqrt[4]{\frac{m\omega}{\pi\hbar}}\frac{1}{\sqrt{\varepsilon}}\exp\biggl\{
    \frac{i}{2\hbar}\frac{m\dot{\varepsilon}}{\varepsilon}x^2+
    \sqrt{\frac{2m\omega}{\hbar}}\frac{x(\alpha-\delta)}{\varepsilon}-
    \frac{1}{2}\frac{\varepsilon^*}{\varepsilon}\alpha^2\notag\\
    &+\alpha\left(\delta^*+\frac{\varepsilon^*}{\varepsilon}\delta\right)-
    \frac{1}{2}\frac{\varepsilon^*}{\varepsilon}\delta^2-\frac{1}{2}{\vert\delta\vert}^2+
    i\int\limits^t_0\im\left(\dot{\delta}\delta^*\right)\,d\tau-\frac{1}{2}{\vert\alpha\vert}^2\biggr\},\\
    \psi_n(x,t) &= \sqrt[4]{\frac{m\omega}{\pi\hbar}}\frac{1}{\sqrt{\varepsilon}}
    \frac{1}{\sqrt{2^nn!}}{\left(\sqrt{\frac{\varepsilon^*}{\varepsilon}}\right)}^n
    \exp\biggl\{\frac{i}{2\hbar}\frac{m\dot{\varepsilon}}{\varepsilon}x^2-
    \sqrt{\frac{2m\omega}{\hbar}}\frac{x\delta}{\varepsilon}\notag\\
    &-\frac{1}{2}\frac{\varepsilon^*}{\varepsilon}\delta^2-\frac{1}{2}{\vert\delta\vert}^2+
    i\int\limits^t_0\im\left(\dot{\delta}\delta^*\right)\,d\tau\biggr\}
    H_n\left(\sqrt{\frac{m\omega}{\hbar}}\frac{x}{\vert\varepsilon\vert}+
    \frac{\delta\varepsilon^*+\delta^*\varepsilon}{\sqrt{2}\vert\varepsilon\vert}\right).
\end{align}
As in general case the coherent state is described by Gaussian wave function. The Fock state
is described by one-dimensional Hermite polynomial. The parameters of the wave functions
depend on the property of the linear integrals of motion of the oscillator. Substituting the
expressions \eqref{e:Xi}--\eqref{e:X0} into \eqref{e:H} and \eqref{e:wn}, we obtain the
tomogram of the coherent state
\begin{equation}
    w_\alpha(X,\mu,\nu,t) = \sqrt{\frac{m\omega}{\pi\hbar}}\frac{1}{\left\vert m\dot{\varepsilon}\nu+\varepsilon\mu\right\vert}
    \exp\biggl\{-\frac{m\omega}{\hbar}\frac{1}{{\left\vert m\dot{\varepsilon}\nu+\varepsilon\mu\right\vert}^2}
    {\left(X-X_0\right)}^2\biggr\}
\end{equation}
and the tomogram of the Fock state
\begin{equation}
    w_n(X,\mu,\nu,t) = w_0(X,\mu,\nu,t)\frac{1}{2^nn!}
    \Biggl|H_n\left(\sqrt{\frac{m\omega}{\hbar}}\frac{X}{\left\vert m\dot{\varepsilon}\nu+\varepsilon\mu\right\vert}+
    \frac{(m\dot{\varepsilon}^*\nu+\varepsilon^*\mu)\delta+(m\dot{\varepsilon}\nu+\varepsilon\mu)\delta^*}
    {\sqrt{2}\left\vert m\dot{\varepsilon}\nu+\varepsilon\mu\right\vert}\right)\Biggr|^2
\end{equation}
of parametric oscillator under consideration.

If driving force is absent then $\delta=0$ and we have the following expressions instead of
previous ones. The wave function of the coherent state is given by the Gaussian
\begin{equation}
    \psi_\alpha(x,t) = \sqrt[4]{\frac{m\omega}{\pi\hbar}}\frac{1}{\sqrt{\varepsilon}}\exp\biggl\{
    \frac{i}{2\hbar}\frac{m\dot{\varepsilon}}{\varepsilon}x^2+
    \sqrt{\frac{2m\omega}{\hbar}}\frac{x\alpha}{\varepsilon}-
    \frac{1}{2}\frac{\varepsilon^*}{\varepsilon}\alpha^2-
    \frac{1}{2}{\vert\alpha\vert}^2\biggr\}.
\end{equation}
The wave function of the Fock state is expressed in terms of Hermite polynomial
\begin{equation}
    \psi_n(x,t) = \sqrt[4]{\frac{m\omega}{\pi\hbar}}\frac{1}{\sqrt{\varepsilon}}
    \frac{1}{\sqrt{2^nn!}}{\left(\sqrt{\frac{\varepsilon^*}{\varepsilon}}\right)}^n
    \exp\biggl\{\frac{i}{2\hbar}\frac{m\dot{\varepsilon}}{\varepsilon}x^2\biggr\}
    H_n\left(\sqrt{\frac{m\omega}{\hbar}}\frac{x}{\vert\varepsilon\vert}\right).
\end{equation}
The tomogram of the coherent state coincides with Gaussian distribution function of the form
\begin{equation}
    w_\alpha(X,\mu,\nu,t) = \sqrt{\frac{m\omega}{\pi\hbar}}\frac{1}{\left\vert m\dot{\varepsilon}\nu+\varepsilon\mu\right\vert}
    \exp\biggl\{-\frac{m\omega}{\hbar}\frac{1}{{\left\vert m\dot{\varepsilon}\nu+\varepsilon\mu\right\vert}^2}
    {\left(X-X_0\right)}^2\biggr\}.
\end{equation}
The tomogram of the Fock state of the parametric oscillator reads
\begin{equation}
    w_n(X,\mu,\nu,t) = w_0(X,\mu,\nu,t)\frac{1}{2^nn!}
    {\left\vert H_n\left(\sqrt{\frac{m\omega}{\hbar}}\frac{X}{\left\vert
    m\dot{\varepsilon}\nu+\varepsilon\mu\right\vert}\right)\right\vert}^2.
\end{equation}
For the case of constant frequency $\omega=\const\not=0$, we have $\varepsilon=e^{i\omega t}$
and the expressions for the wave functions of the coherent and Fock states and the tomograms
of these states are given by simple formulas. The wave function of the coherent state of the
harmonic oscillator in position representation reads
\begin{equation}
    \psi_\alpha(x,t) = \sqrt[4]{\frac{m\omega}{\pi\hbar}}\exp\biggl\{
    -\frac{m\omega}{2\hbar}x^2+
    \sqrt{\frac{2m\omega}{\hbar}}e^{-i\omega t}x\alpha-
    \frac{1}{2}e^{-2i\omega t}\alpha^2-
    \frac{1}{2}{\vert\alpha\vert}^2-\frac{1}{2}i\omega
    t\biggr\}.
\end{equation}
The wave function of the stationary state of the harmonic oscillator describing for the
one-mode electromagnetic field the $n$-photon state in the position representation has the
standard form
\begin{equation}
    \psi_n(x,t) = \sqrt[4]{\frac{m\omega}{\pi\hbar}}
    \frac{1}{\sqrt{2^nn!}}e^{-i\omega t\left(n+\frac{1}{2}\right)}
    \exp\biggl\{-\frac{m\omega}{2\hbar}x^2\biggr\}
    H_n\left(\sqrt{\frac{m\omega}{\hbar}}x\right).
\end{equation}
The tomogram of the coherent state of the harmonic oscillator reads
\begin{equation}
    w_\alpha(X,\mu,\nu,t) = \sqrt{\frac{m\omega}{\pi\hbar}}\frac{1}{\sqrt{\mu^2+m^2\omega^2\nu^2}}
    \exp\biggl\{-\frac{m\omega}{\hbar}\frac{1}{\mu^2+m^2\omega^2\nu^2}{\left(X-X_0\right)}^2\biggr\}.
\end{equation}
The tomogram of the $n$-th excited state of the harmonic oscillator takes the form
\begin{equation}
    w_n(X,\mu,\nu,t) = w_0(X,\mu,\nu,t)\frac{1}{2^nn!}
    {\left\vert H_n\left(\sqrt{\frac{m\omega}{\hbar}}\frac{X}{\sqrt{\mu^2+m^2\omega^2\nu^2}}\right)\right\vert}^2
\end{equation}
Thus we obtained explicit forms of tomograms for the specific quantum states of parametric and
harmonic oscillators.

\subsection{Charged Particle in Nonstationary Electric Field}

Important partial case of quadratic system is the problem of
motion of a charged particle in a homogeneous time-dependent
electric field. The Hamiltionian of the particle reads
\begin{equation}
    \widehat{H}(t) = \frac{\widehat{p}^2}{2m} + F(t)\widehat{x},
\end{equation}
where $F(t)$ is nonstationary electric field. Here the mtarices and vectors determining the
generic Hamiltonian of a quadratic system \eqref{e:quadricH} are reduced to the numbers
$B_{pp} = 1/m$, $B_{px} = B_{xp} = B_{xx} = 0$, $c_p = 0$, $c_x(t) = F(t)$ and for matrices
$\Lambda_p$, $\Lambda_x$ and vector $\delta$ we have
\begin{align} \label{e:15}
    \Lambda_x(t) &= A_x, \\
    \Lambda_p(t) &= -\frac{1}{m}A_xt+A_p, \\
    \delta(t) & = \int\limits^t_0\left(A_p-\frac{A_x}{m}t\right)F(\tau)\,d\tau.
\end{align}
Using  expressions \eqref{e:15} one can obtain the wave function of the coherent state of the
charged particle
\begin{align}
\begin{split}
    \psi_\alpha(x,t) &= \frac{1}{\sqrt[4]{2\pi\hbar^2}}\frac{1}{\sqrt{A_p-\frac{A_x}{m}t}}
    \exp\biggl\{-\frac{i}{2\hbar}\frac{A_x}{A_p-\frac{A_x}{m}t}x^2+
    \frac{i}{\hbar}\frac{x(\alpha-\delta)}{A_p-\frac{A_x}{m}t}+\frac{1}{2}
    \frac{A^*_p-\frac{A^*_x}{m}t}{A_p-\frac{A_x}{m}t}\alpha^2\\
    &+\alpha\left(\delta^*-\frac{A^*_p-\frac{A^*_x}{m}t}{A_p-\frac{A_x}{m}t}\delta\right)+
    \frac{1}{2}\frac{A^*_p-\frac{A^*_x}{m}t}{A_p-\frac{A_x}{m}t}\delta^2-
    \frac{1}{2}{\vert\delta\vert}^2+i\int\limits^t_0\im\left(\dot{\delta}\delta^*\right)\,d\tau-
    \frac{1}{2}{\vert\alpha\vert}^2\biggr\},
\end{split}
\end{align}
and the wave function of the Fock state of the particle
\begin{align}
\begin{split}
    \psi_n(x,t) &= \frac{1}{\sqrt[4]{2\pi\hbar^2}}\frac{1}{\sqrt{A_p-\frac{A_x}{m}t}}\frac{1}{\sqrt{2^nn!}}
    {\left(\sqrt{-\frac{A^*_p-\frac{A^*_x}{m}t}{A_p-\frac{A_x}{m}t}}\;\right)}^2
    \exp\biggl\{-\frac{i}{2\hbar}\frac{A_x}{A_p-\frac{A_x}{m}t}x^2-
    \frac{i}{\hbar}\frac{x\delta}{A_p-\frac{A_x}{m}t}\biggr.\\
    \biggl.&+\frac{1}{2}\frac{A^*_p-\frac{A^*_x}{m}t}{A_p-\frac{A_x}{m}t}\delta^2-
    \frac{1}{2}{\vert\delta\vert}^2 +
    i\int\limits^t_0\im\left(\dot{\delta}\delta^*\right)\,d\tau\biggr\}\\
    &\times H_n\left(\frac{1}{\hbar}\frac{x}{\sqrt{2}\left\vert A_p-\frac{A_x}{m}t\right\vert}+
    i\frac{\left(A^*_p-\frac{A^*_x}{m}t\right)\delta-\left(A_p-\frac{A_x}{m}t\right)\delta^*}
    {\sqrt{2}\left\vert A_p-\frac{A_x}{m}t\right\vert}\right).
\end{split}
\end{align}
These states are nonstationary normalized states of the charge in the electric field.
Tomograms of coherent and Fock states are given by the formulas \eqref{e:H} and
\eqref{e:walph1} in which the matrix $\Xi$, the dispersion matrix $\Sigma$ and vector $X_0$
take the following values:
\begin{align}
    \Xi &= iA_x\nu - i\left(A_p-\frac{A_x}{m}t\right)\mu, \\
    \Sigma &= \hbar^2{\left\vert A_x\nu-\left(A_p-\frac{A_x}{m}t\right)\mu\right\vert}^2, \\
    X_0 &= \hbar\left({\left(iA_x\nu - i\left(A_p-\frac{A_x}{m}t\right)\mu\right)}^*(\alpha-\delta) +
           \left(iA_x\nu - i\left(A_p-\frac{A_x}{m}t\right)\mu\right)(\alpha^*-\delta^*)\right).
\end{align}
The parameters $A_x$ and $A_p$ determine the degree of squeezing in the coherent state of the
charge in the electric field.

\section{Conclusion}
We have shown that for multimode systems with Hamiltonian which is quadratic in position and
momentum operators the quantum states can be described by symplectic tomograms. For coherent
states these tomograms are the multivariable Gaussian distribution functions. For the Fock
states the tomograms are expressed in terms of multivariable Hermite polynomials. New results
of our work are the explicit formulas for tomograms of coherent and Fock states of multimode
nonstationary quadratic systems and new relations for multivariable Hermite polynomials. Since
many physical systems including charge moving in electric and magnetic fields and photons in a
resonator with moving boundaries are described by quadratic Hamiltonians the results of this
work can be applied to consider the evolution of quantum states of these physical systems in
the tomographic probability representation. The mathematical formalism of these quadratic
systems is based on properties of multivariable Hermite polynomials (\cite{qso-9-381},
\cite{jmp-35-4277}, \cite{jp-a-27-6191}, \cite{jp-a-34-6185}). Due to this the new formulas
for multivariable Hermite polynomials obtained in the tomographic representation can be used
in analysis of the different physical systems with quadratic Hamiltonians. It is worthy to
consider some other particular cases of multimode quadratic systems. We hope to do this in
forthcoming paper.

\appendix

\section{Properties of symplectic $\Lambda$-matrices}

Using definition of operators $\widehat{{\boldsymbol{a}}}$ and
$\widehat{{\boldsymbol{a}}}^+$ we obtain the following properties
of matrices $A_p$ and $A_x$:
\begin{align}
    A_xA^T_p - A_pA^T_x &= 0, \label{e:nA1}\\
    A_xA^+_p - A_pA^+_x &= -\frac{i}{\hbar}E_N. \label{e:nA2}
\end{align}
Using these two properties, one can prove that the matrices $A_p$ and $A_x$ determine the real
symplectic transform of position and momentum, i.e.
\begin{enumerate}\label{e:LambdaM}
    \renewcommand{\theenumi}{(\roman{enumi})}
    \renewcommand{\labelenumi}{\theenumi}

    \item
    Matrices $A_p$ and $A_x$ are non-singular,

    \item
    $A^T_pA^*_p = A^+_pA_p,\quad A^T_xA^*_x=A^+_xA_x$,

    \item
    $A^+_xA_p-A^T_xA^*_p = A^T_pA^*_x-A^+_pA_x =
    \displaystyle\frac{i}{\hbar}E_N$,
\end{enumerate}

We prove first the property (i). Suppose that matrix $A_p$ is singular. Then the transposed
matrix $A^T_p$ is also singular and there exists a nonzero vector ${\boldsymbol{z}}$ such that
$A^+_p{\boldsymbol{z}} = 0$. From \eqref{e:nA2} we have the chain of the relations for the
scalar products
\begin{equation}
    0 \not= \frac{i}{\hbar}{\Vert\boldsymbol{z}\Vert}^2 =
    \frac{i}{\hbar}\left(\boldsymbol{z},\boldsymbol{z}\right) =
    \left(\boldsymbol{z},A_xA^+_p\boldsymbol{z}\right) -
    \left(\boldsymbol{z},A_pA^+_x\boldsymbol{z}\right) =
    -\left(A^+_p\boldsymbol{z},A^+_x\boldsymbol{z}\right) = 0
\end{equation}
These contradictory relations show the nonsingularity of the matrix $A_p$. The proof that the
matrix $A_x$ is also nonsingular is analogous.

Let us prove the property (ii). In view of nonsingularity of the matrices $A_p$ and $A_x$ we
can write \eqref{e:nA2} in form
\begin{equation}\label{e:xp}
    A^+_x-A^{-1}_pA_xA^+_p = \frac{i}{\hbar}A^{-1}_p.
\end{equation}
From \eqref{e:nA1} it follows that
\begin{equation}\label{e:-1}
    A^{-1}_pA_x = A^T_x{A^T_p}^{-1}.
\end{equation}
Also we can obtain that
\begin{equation}
    A^+_x = {A^*_p}^{-1}A^*_xA^+_p.
\end{equation}
Substituting this expression for $A^+_x$ and the equality \eqref{e:-1} in \eqref{e:xp} we get
\begin{equation}
    {A^*_p}^{-1}A^*_xA^+_p-A^T_x{A^T_p}^{-1}A^+_p =
    \frac{i}{\hbar}A^{-1}_p.
\end{equation}
This formula can be rewritten in the form
\begin{equation}
    {A^*_p}^{-1}\left(A^*_xA^T_p-A^*_pA^T_x\right){A^T_p}^{-1}A^+_p
    = \frac{i}{\hbar}A^{-1}_p.
\end{equation}
Using hermitian conjugate of \eqref{e:nA2} we see that the above equation gives the relation
\begin{equation}
    {A^*_p}^{-1}{A^T_p}^{-1}A^+_p = A^{-1}_p
\end{equation}
or
\begin{equation}
    A^+_pA_p = A^T_pA^*_p.
\end{equation}
The second equality in (ii) can be proved analogously.

Let us prove the property (iii). Multipling \eqref{e:nA2} by $A^+_x$ from the left side we get
\begin{equation}
    A^+_xA_xA^+_p - A^+_xA_pA^+_x = -\frac{i}{\hbar}A^+_x.
\end{equation}
Using the property (ii) we get the relation
\begin{equation}
    A^+_xA_pA^+_x - A^T_xA^*_xA^+_p = \frac{i}{\hbar}A^+_x,
\end{equation}
or using \eqref{e:nA1} we get
\begin{equation}
    A^+_xA_pA^+_x - A^T_xA^*_pA^+_x = \frac{i}{\hbar}A^+_x.
\end{equation}
Since the matrix $A_x$ and consequently the matrix $A^+_x$ are nonsingular we obtain the
relation (iii).

Calculating the time derivative of the product of the matrices
$\Lambda(t)\Sigma_{2N}\Lambda^T(t)$ we get the equality
\begin{equation}
    \frac{d}{dt}\Bigl(\Lambda(t)\Sigma_{2N}\Lambda^T(t)\Bigr) =
    \dot{\Lambda(t)}\Sigma_{2N}\Lambda^T(t)+\Lambda(t)\Sigma_{2N}{\dot{\Lambda}}^T(t) =
    0.
\end{equation}
Taking into account the initial condition for matrix $\Lambda(t)$ \eqref{e:ini}, we obtain
\begin{equation}
    \Lambda(t)\Sigma_{2N}\Lambda^T(t) = \Sigma_{2N}.
\end{equation}
Making the same procedure with the product of matrices $\Omega(t)\Sigma_{2N}\Omega^T(t)$ we
obtain
\begin{equation}\label{e:omega}
   \Omega(t)\Sigma_{2N}\Omega^T(t) = \frac{i}{\hbar}E_{2N}.
\end{equation}
Rewriting \eqref{e:omega} in more detail, we have
\begin{align}
    \Lambda_x\Lambda^T_p - \Lambda_p\Lambda^T_x &= 0, \\
    \Lambda_x\Lambda^+_p - \Lambda_p\Lambda^+_x &=
    -\frac{i}{\hbar}E_N.
\end{align}
These equations are similar to \eqref{e:nA1} and \eqref{e:nA2} and therefore the matrices
$\Lambda_p$ and $\Lambda_x$ possess the same properties as $A_p$ and $A_x$, i.e.,
\begin{enumerate}
    \renewcommand{\theenumi}{(\roman{enumi})}
    \renewcommand{\labelenumi}{\theenumi}

    \item
    Matrices $\Lambda_p$ and $\Lambda_x$ are non-singular,

    \item
    $\Lambda^T_p\Lambda^*_p = \Lambda^+_p\Lambda_p,\quad
    \Lambda^T_x\Lambda^*_x=\Lambda^+_x\Lambda_x$,

    \item
    $\Lambda^+_x\Lambda_p-\Lambda^T_x\Lambda^*_p = \Lambda^T_p\Lambda^*_x-\Lambda^+_p\Lambda_x =
    \displaystyle\frac{i}{\hbar}E_N$.
\end{enumerate}

\section{Calculation of the Matrix Exponent}
In this appendix we calculate the exponent
\begin{equation}\label{e:180}
    \exp\left\{\begin{Vmatrix}0 & -B_{pp} \\ \\ B_{xx} &
    0\end{Vmatrix}t\right\} =
    \sum\limits^{+\infty}_{n=0}\frac{1}{n!}
    {\begin{Vmatrix}0 & -B_{pp} \\ \\ B_{xx} &
    0\end{Vmatrix}}^nt^n =
    \begin{Vmatrix}E_{pp}(t) & E_{px}(t) \\ \\ E_{xp}(t) & E_{xx}(t)\end{Vmatrix}.
\end{equation}
To calculate this exponent it is worthy to note that
\begin{align}
\begin{split}
    {\begin{Vmatrix}0 & -B_{pp} \\ \\ B_{xx} & 0\end{Vmatrix}}^{2k} &=
    \begin{Vmatrix}-{\left(B_{pp}B_{xx}\right)}^k & 0 \\  \\ 0 &
    -{\left(B_{xx}B_{pp}\right)}^k\end{Vmatrix}, \\
    {\begin{Vmatrix}0 & -B_{pp} \\ \\ B_{xx} & 0\end{Vmatrix}}^{2k+1} &=
    \begin{Vmatrix} 0 & \left(-B_{pp}\right){\left(-B_{xx}B_{pp}\right)}^k \\ \\ \left(-B_{xx}\right){\left(-B_{pp}B_{xx}\right)}^k & 0
    \end{Vmatrix}.
\end{split}
\end{align}
In view of this we get
\begin{align} \label{e:11}
\begin{split}
     E_{pp}(t) &= E_N-\frac{1}{2!}B_{pp}B_{xx}t^2+\frac{1}{4!}{\left(B_{pp}B_{xx}\right)}^2t^4-\ldots =
    \cos\left(\sqrt{B_{pp}B_{xx}}\;t\right), \\
     E_{px}(t) &= -B_{pp}t+\frac{1}{3!}B_{pp}B_{xx}B_{pp}t^3-\frac{1}{5!}B_{pp}{\left(B_{xx}B_{pp}\right)}^2t^5+\ldots
        =-\sqrt{B_{pp}B^{-1}_{xx}}\sin\left(\sqrt{B_{xx}B_{pp}}\;t\right), \\
    E_{xp}(t) &= B_{xx}t-\frac{1}{3!}B_{xx}B_{pp}B_{xx}t^3+\frac{1}{5!}B_{xx}{\left(B_{pp}B_{xx}\right)}^2t^5+\ldots
        =-\sqrt{B_{xx}B^{-1}_{pp}}\sin\left(\sqrt{B_{pp}B_{xx}}\;t\right), \\
    E_{xx}(t) &= E_N-\frac{1}{2!}B_{xx}B_{pp}t^2+\frac{1}{4!}{\left(B_{xx}B_{pp}\right)}^2t^4-\ldots =
    \cos\left(\sqrt{B_{xx}B_{pp}}\;t\right).
\end{split}
\end{align}
Thus for the matrix exponent \eqref{e:180} we have the expression
\begin{equation}
    \exp\left\{\begin{Vmatrix}0 & -B_{pp} \\ \\ B_{xx} &
    0\end{Vmatrix}t\right\} = \begin{Vmatrix}\cos\left(\sqrt{B_{pp}B_{xx}}\;t\right) & -\sqrt{B_{pp}B^{-1}_{xx}}\sin\left(\sqrt{B_{xx}B_{pp}}\;t\right) \\ \\
                   \sqrt{B_{xx}B^{-1}_{pp}}\sin\left(\sqrt{B_{pp}B_{xx}}\;t\right) &
                   \cos\left(\sqrt{B_{xx}B_{pp}}\;t\right)\end{Vmatrix}.
\end{equation}
Solution \eqref{e:o} is
\begin{equation}
    \begin{Vmatrix}\Lambda^T_p(t) \\  \\ \Lambda^T_x(t)\end{Vmatrix} =
    \begin{Vmatrix}\cos\left(\sqrt{B_{pp}B_{xx}}\;t\right) & -\sqrt{B_{pp}B^{-1}_{xx}}\sin\left(\sqrt{B_{xx}B_{pp}}\;t\right) \\ \\
                   \sqrt{B_{xx}B^{-1}_{pp}}\sin\left(\sqrt{B_{pp}B_{xx}}\;t\right) & \cos\left(\sqrt{B_{xx}B_{pp}}\;t\right)\end{Vmatrix}
    \begin{Vmatrix}A^T_p \\  \\ A^T_x\end{Vmatrix},
\end{equation}
or
\begin{align}
    \Lambda_p(t) &= A_p\cos\left(\sqrt{B_{pp}B_{xx}}\;t\right) - A_x\sqrt{B_{pp}B^{-1}_{xx}}\sin\left(\sqrt{B_{xx}B_{pp}}\;t\right), \\
    \Lambda_x(t) &= A_p\sqrt{B_{xx}B^{-1}_{pp}}\sin\left(\sqrt{B_{pp}B_{xx}}\;t\right) +
    A_x\cos\left(\sqrt{B_{xx}B_{pp}}\;t\right).
\end{align}
If we take $A_p$ and $A_x$ as
\begin{equation}
    A_p =
    \frac{i}{\sqrt{2\hbar}}\sqrt[4]{B^{-1}_{pp}B_{xx}},\qquad\qquad
    A_x = \frac{i}{\sqrt{2\hbar}}\sqrt[4]{B_{pp}B^{-1}_{xx}},
\end{equation}
we have
\begin{align}
    \Lambda_p(t) &= \frac{i}{\sqrt{2\hbar}}\sqrt[4]{B_{pp}B^{-1}_{xx}}\;\exp\left\{i\sqrt{B_{pp}B_{xx}}t\right\}, \\
    \Lambda_x(t) &=
    \frac{1}{\sqrt{2\hbar}}\sqrt[4]{B^{-1}_{pp}B_{xx}}\;\exp\left\{i\sqrt{B_{pp}B_{xx}}t\right\}.
\end{align}
These expressions were used in the main text.

\end{document}